\documentclass[preprint,12pt]{elsarticle}
\usepackage{amssymb}
\usepackage[utf8]{inputenc}   
\graphicspath{ {./figures/} }
\usepackage{hyperref}
\usepackage{floatrow}
\usepackage{verbatim} 
\usepackage{amsfonts}
\usepackage{booktabs}
\usepackage{multirow}
\usepackage{threeparttable}
\usepackage{tabularx}
\usepackage{makecell}
\usepackage{hyperref}
\usepackage{lineno}
\usepackage{graphicx} 
\usepackage{subfig}      
\usepackage[ruled]{algorithm2e} 
\usepackage{amsmath}
\usepackage{float}
\usepackage{setspace}
\usepackage{amsthm}

\floatstyle{plain}  
\restylefloat{figure}
\restylefloat{figure}
\restylefloat{table}

\journal{Engineering Applications of Artificial Intelligence}

\begin{document}


\floatsetup[table]{capposition=top}

\begin{frontmatter}

\title{Physics-Aware Compilation for Parallel Quantum Circuit Execution on Neutral Atom Arrays}

\author[lable1]{Geng Chen}
\author[lable1]{Guowu Yang}
\author[lable2]{Wenjie Sun}
\author[lable3]{Lianhui Yu}
\author[lable4]{Guangwei Deng}
\author[lable6, lable7]{Desheng Zheng}
\author[lable7]{XiaoYu Li$^\dagger$}

\affiliation[lable1]{organization={School of Computer Science and Engineering, University of Electronic Science and Technology of China},
	city={Cheng Du},
	postcode={610054}, 
	state={Si Chuan},
	country={China}}
\affiliation[lable2]{organization={School of Electronic Science and Engineering, University of Electronic Science and Technology of China},
	city={Cheng Du},
	postcode={610054}, 
	state={Si Chuan},
	country={China}}
\affiliation[lable3]{organization={School of Physics, University of Electronic Science and Technology of China},
	city={Cheng Du},
	postcode={610054}, 
	state={Si Chuan},
	country={China}}
\affiliation[lable4]{organization={Institute of Fundamental and Frontier Sciences, University of Electronic Science and Technology of China},
	city={Cheng Du},
	postcode={610054}, 
	state={Si Chuan},
	country={China}}
\affiliation[lable5]{organization={Kash Institute of Electronics and Information Industry},
	city={Kashgar},
	postcode={844199}, 
	state={XinJiang},
	country={China}}
\affiliation[lable6]{organization={School of Computer Science and Software Engineering, Southwest Petroleum University},
	city={Cheng Du},
	postcode={610599}, 
	state={Si Chuan},
	country={China}}
\affiliation[lable7]{organization={School of Information and Software Engineering, University of Electronic Science and Technology of China},
	city={Cheng Du},
	postcode={610054}, 
	state={Si Chuan},
	country={China}}

\begin{abstract}
Neutral atom quantum computers are one of the most promising quantum architectures, offering advantages in scalability, dynamic reconfigurability, and potential for large-scale implementations. These characteristics create unique compilation challenges, especially regarding compilation efficiency while adapting to hardware flexibility. However, existing methods encounter significant performance bottlenecks at scale, hindering practical applications. We propose $\textbf{P}$hysics-$\textbf{A}$ware $\textbf{C}$ompilation (PAC), a method that improves compilation efficiency while preserving the inherent flexibility of neutral atom systems. PAC introduces physics-aware hardware plane partitioning that strategically allocates hardware resources based on physical device characteristics like AOD and SLM trap properties and qubit mobility constraints. Additionally, it implements parallel quantum circuit division with an improved Kernighan-Lin algorithm that enables simultaneous execution across independent regions while maintaining circuit fidelity. Our experimental evaluation compares PAC with state-of-the-art methods across increasingly larger array sizes ranging from $16\times16$ to $64\times64$ qubits. Results demonstrate that PAC achieves up to 78.5× speedup on $16\times16$ arrays while maintaining comparable circuit quality. PAC$'$s compilation efficiency advantage increases with system scale, demonstrating scalability for practical quantum applications on larger arrays. PAC explores a viable path for practical applications of neutral atom quantum computers by effectively addressing the tension between compilation efficiency and hardware flexibility. \href{https://github.com/StillwaterQ/PAC}{Code} for PAC is available.
\end{abstract}

\begin{graphicalabstract}
\begin{figure}[H]
	\hspace{0cm}
	\includegraphics[width=1\textwidth, trim=1.8cm 7.5cm 1.8cm 0cm, clip]{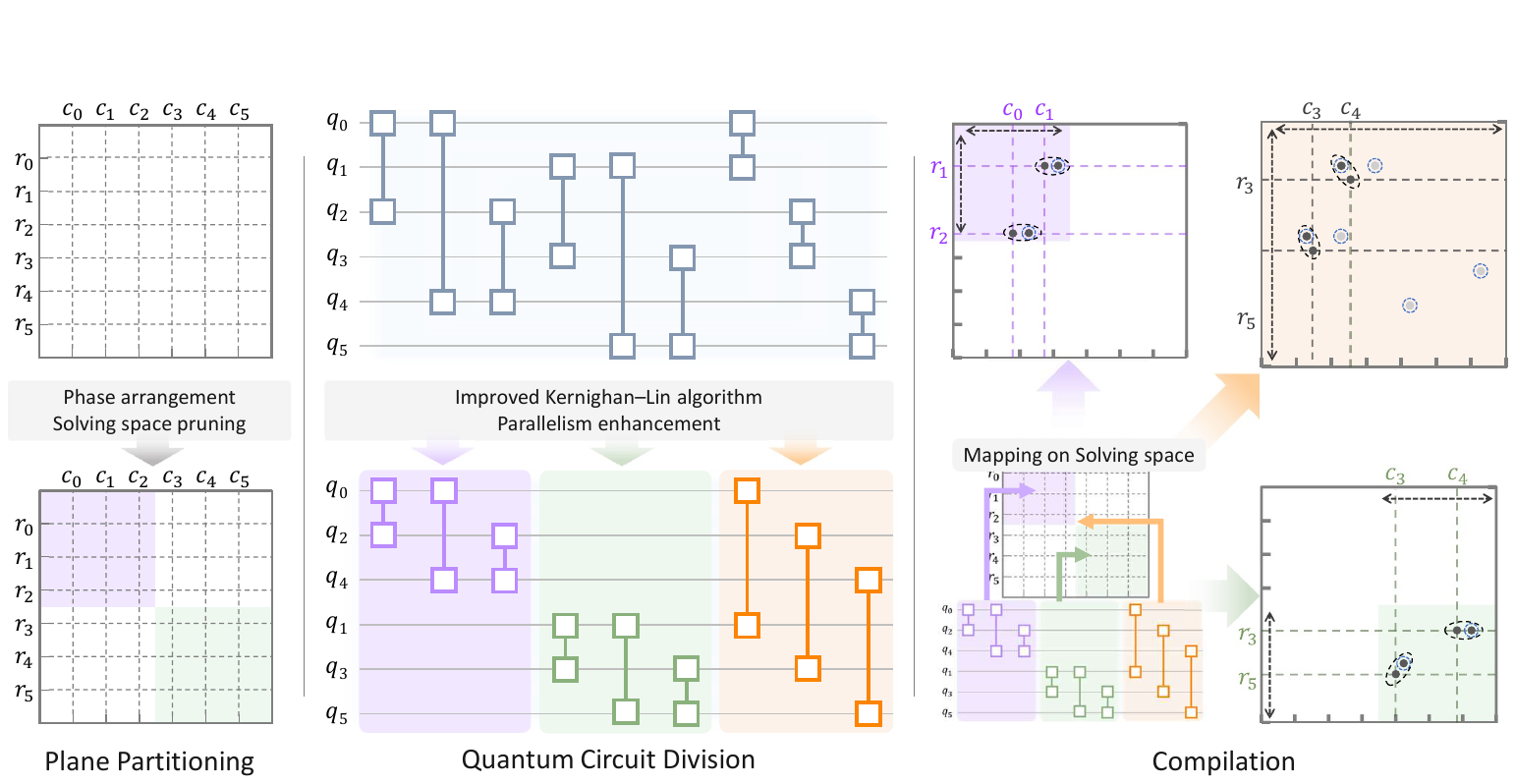}
\end{figure}
\end{graphicalabstract}

\begin{highlights}
	\item Presents \textbf{P}hysics-\textbf{A}ware \textbf{C}ompilation balancing efficiency with flexibility in neutral atom quantum systems.
	\item Achieves parallel quantum circuit execution through hardware plane partitioning and improved Kernighan-Lin algorithm.
	\item Delivers up to $202.81\times$ speedup while maintaining comparable circuit quality across various array sizes.
\end{highlights}


\end{frontmatter}

\section{Introduction}
\label{introduction}

Quantum computing has powerful parallel computing capabilities and has the potential to solve optimization problems\cite{Harrigan2021}, perform chemical simulations\cite{GAQC2020}, solve linear systems of equations\cite{Cerezo2021}, etc. In order to better utilize the advantages of quantum computing, a variety of hardware technology routes have been proposed to achieve real quantum computing, including superconducting\cite{Arute2019}, quantum dot\cite{Thomas2025}, atom arrays\cite{Bluvstein2022}, etc.  Among these platforms, neutral atom quantum computing has attracted growing interest due to its advantages in large-scale quantum computation and inherent reconfigurability, offering flexibility in qubit arrangement and interaction.

Quantum compilation is an essential process for deploying quantum circuits onto quantum computers. It involves transforming a quantum circuit into an executable sequence of operations that can be implemented on quantum hardware to achieve the intended functionality. The quantum compilation approach varies significantly depending on the underlying quantum hardware architecture. Currently, most quantum compilation research focuses on superconducting quantum computers, encompassing solver-based methods \cite{Tan2020,Tan2021,Lin2023}, heuristic approaches \cite{Li2019,Niu2020,Huang2024}, and other techniques.
Recently, neutral atom quantum computers have attracted growing interest due to their advantages in large-scale quantum computation \cite{Manetsch2024} and their inherent reconfigurability \cite{Bluvstein2024}. Unlike superconducting quantum computers, the hardware implementation of quantum operations in neutral atom systems is not strictly constrained by swap gates. By trapping atoms that serve as qubit carriers in an optical tweezer array \cite{Bluvstein2022}, individual qubits can be flexibly transported and brought into proximity with one another. This mobility facilitates the formation of entanglement and enables an extended quantum lifetime \cite{Henriet2020}.

However, this flexibility introduces a critical paradox in quantum compilation: existing methods either optimize compilation efficiency at the expense of hardware flexibility, or preserve flexibility while suffering performance bottlenecks. This tension becomes particularly acute as system sizes increase, with compilation times for neutral atom platforms reaching hours for moderately-sized circuits. Such inefficiency severely constrains the practical utility of neutral atom quantum computers despite their theoretical advantages.

As an attempt to address this key contradiction, a series of fruitful efforts have been put forward. Baker\cite{Baker2021}introduced a layered compilation approach for quantum circuits. However, the inter-layer scheduling introduced additional execution time. In 2023, Li et al. \cite{Li2023} proposed a greedy algorithm incorporating Monte Carlo tree search to optimize circuit compilation. Patel et al. presented GRAPHINE \cite{Patel2023}, which utilizes an annealing-based approach for qubit position arrangement. As a mainstream solution, reconfigurable quantum processors have been developed \cite{Bluvstein2022}. These architectures enable flexible atomic repositioning, facilitating entanglement between different qubits and achieving enhanced connectivity, programmability, and parallelism. Leveraging this architecture, Tan et al. (2022) proposed the first solver-based compilation method for neutral atoms \cite{Tan2022}. In 2024, Tan et al. introduced DPQA\cite{Tan2024}, which integrates the advantages of greedy algorithms and solver-based approaches to improve both computational efficiency and scalability. In the same year, Ludmir et al. proposed PARALLAX \cite{Ludmir2024}, an extension of the GRAPHINE method that enhances circuit execution fidelity.

While these approaches have shown promise, they fail to effectively address the fundamental tension between leveraging the platform's inherent parallelization potential and maintaining its flexible reconfigurability. As a result, existing compilation methods for neutral atom quantum computers fail to fully exploit their inherent parallelization potential and the resulting improvements in efficiency. These approaches treat the compilation problem as primarily computational rather than recognizing it as inherently problems in the physical scene —a perspective shift that could potentially resolve the parallelism-flexibility paradox.

As a response to this motivation, we propose a physics-aware compilation (PAC) approach for neutral atom quantum computers. PAC recognizes that the unique physical characteristics of neutral atom hardware can guide compilation decisions. PAC introduces a Physics-Aware Hardware Plane Partitioning method that strategically allocates hardware resources based on the physical characteristics of the device. This hardware-aware approach effectively prunes the solution space at the hardware level while enabling optimal task allocation. Furthermore, PAC implements Parallel Quantum Circuit Division with an improved Kernighan-Lin algorithm that divides qubits of the target circuit into two communities with tight internal connections and fewer cross-community relationships. This approach enables simultaneous execution across independent regions while maintaining circuit fidelity, enhancing compilation parallelism and ultimately improving overall efficiency. Together, these techniques address the critical tension between compilation speed and hardware flexibility. The main contribution of this paper is:

\begin{enumerate}
	\item This work focuses on the critical tension between compilation efficiency and hardware flexibility in neutral atom quantum computing. By analyzing physical properties of neutral atom systems, a physics-aware compilation approach is proposed that significantly improves efficiency while preserving the platform's inherent reconfigurability.
	
	\item A two-component framework is developed, consisting of Hardware Plane Partitioning for physics-guided resource allocation and Quantum Circuit Division. This approach effectively transforms complex global compilation challenges into manageable local optimization problems while respecting hardware constraints.
	
	\item Extensive experimental evaluation across multiple array scales demonstrates PAC$`$s performance on diverse benchmark circuits. PAC achieves up to $78.5\times$ speedup compared to SOTA method. PAC successfully compiles all test circuits on arrays up to $64\times64$ qubits,  circuit depth across configurations comparable to existing methods while dramatically reducing compilation time.
	
\end{enumerate}

\section{Related Works}
\label{Related Works}

\subsection{Reconfigurable Neutral Atom Computing}
\label{Reconfigurable Neutral Atom Computing}

Neutral atom quantum computing has attracted growing interest due to its advantages in large-scale quantum computation and inherent reconfigurability \cite{Manetsch2024}. In reconfigurable neutral atom quantum computers, qubits are trapped within optical tweezers using two primary trap types: spatial light modulators (SLM) and crossed two-dimensional acousto-optic deflectors (AOD) \cite{Bluvstein2024}. The key distinction between these trap types lies in their mobility—SLM-generated traps remain fixed in position, whereas AOD-generated traps can be dynamically moved in space. When two atoms are brought sufficiently close, they can be illuminated by a Rydberg laser, enabling the execution of entangling gates. Intuitively, AOD traps function as "transporters" that shuttle atoms near other atoms with which they need to form entangling gates, thereby facilitating multi-qubit gate operations.

\begin{figure}[t]
	\centering
	\includegraphics[width=0.8\columnwidth, trim=10cm 8cm 8.5cm 8.5cm, clip]{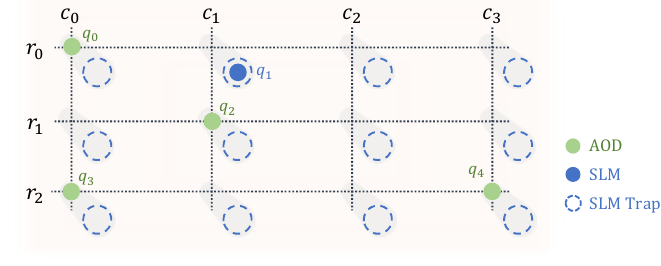}
	\caption{A reconfigurable atom array with AOD and SLM traps.}
	\label{fig:aod-slm}
\end{figure}

As shown in Fig.~\ref{fig:aod-slm}, in a reconfigurable neutral atom array with both AOD and SLM traps, qubits can reside in two types of traps simultaneously. The AOD trap forms a rectangular array with current implementations achieving $16 \times 16$ configurations \cite{Singh2022}, while the SLM trap can generate trapping sites at arbitrary positions within the plane. This arrangement enables flexible atomic repositioning, facilitating entanglement between different qubits and achieving enhanced connectivity and programmability.

However, the flexibility and reconfigurability that make neutral atom platforms attractive also create unique compilation challenges. The compilation process must account for hardware-specific constraints as follow:

\begin{itemize}
	\item \textbf{AOD crossing constraint:} no two columns or rows within an AOD array may intersect.
	\item \textbf{Parallel execution constraint:} when two qubits undergo a two-qubit gate operation, no third qubit can be present within a certain vicinity.
	\item \textbf{Rydberg interaction constraint:} two qubits must be positioned within the Rydberg blockade radius $r_b$ to enable entangling gate operations between them.
\end{itemize}

All neutral atom quantum computers must satisfy these hardware-specific requirements to function properly, representing universal challenges across this computing paradigm. As the technology advances, the solution space becomes dramatically expansive, creating a tension between compilation efficiency and hardware flexibility that must be resolved to fully exploit the potential of neutral atom quantum computing. 

\subsection{Compilation for Reconfigurable Neutral Atom Arrays}

The physical constraints of reconfigurable neutral atom systems, particularly the AOD crossing constraint and parallel execution constraint, complicate the quantum compilation process. These constraints transform the compilation problem from a purely computational challenge into one that must deeply integrate hardware physics considerations. As noted in ~\ref{Reconfigurable Neutral Atom Computing}, the AOD crossing constraint prevents certain qubit movements, as they would require crossing AOD columns or rows. Meanwhile, the parallel execution constraint limits which gates can be executed simultaneously, as qubits involved in gate operations must maintain specific separation distances.

Fig.~\ref{fig:aod-slm} illustrates a reconfigurable atom array architecture with both AOD and SLM traps represented on a coordinate grid. The horizontal axis (columns $c_0$, $c_1$, $c_2$, $c_3$) and vertical axis (rows $r_0$, $r_1$, $r_2$) form a coordinate system for precise qubit positioning. In this representation, AOD traps are depicted at grid intersections where vertical and horizontal acousto-optic deflectors meet, shown as solid circles. The SLM trap, on the other hand, can generate trapping sitesat arbitrary positions within the plane. However to enhance parallelism, it is typically structured similarly to the AOD array, forming a rectangular grid \cite{Tan2022}. Each quantum bit is assigned to a specific coordinate position, for example, qubit $q_0$ is positioned at the intersection of $(r_0, c_0)$ in an AOD trap, while qubit $q_1$ is shown in an SLM trap. Here's an example to show how reconfigurable neutral atom array works.

\begin{figure}[t]
	\centering
	\includegraphics[width=0.8\columnwidth, trim=8.5cm 8.2cm 8.5cm 8.2cm, clip]{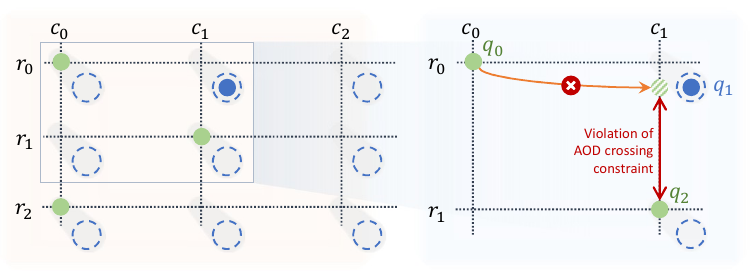}
	\caption{Illustration of AOD crossing constraint violation.}
	\label{Violation-no-cross}
\end{figure}

In a reconfigurable neutral atom computer, qubits can move with AOD traps. But their movement is subject to constraints and the most critical being the AOD crossing constraint. Fig.~\ref{Violation-no-cross} illustrates this constraint, which extracts a section of Fig.~\ref{fig:aod-slm}, highlighting two vertically and two horizontally oriented acousto-optic deflectors. If qubit $q_0$ at $(r_0, c_0)$ needs to establish entanglement with qubit $q_1$ in the SLM trap, directly moving $q_0$ to $q_1$ is prohibited, as it would cause column $c_0$ to cross over column $c_1$, violating the AOD crossing constraint.

\begin{figure}[t]
	\centering
	\includegraphics[width=0.9\columnwidth, trim=8.5cm 8.6cm 8.5cm 8.6cm, clip]{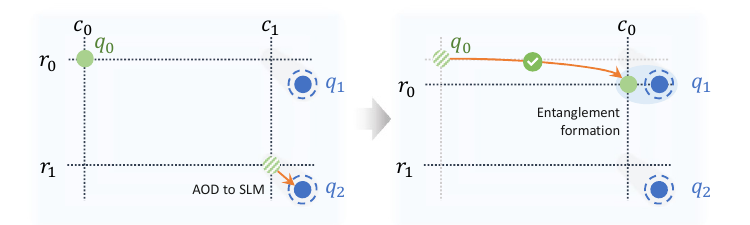}
	\caption{Qubit transition from AOD to SLM trap for entanglement formation.}
	\label{non-crossing}
\end{figure}

To address this issue, one approach is illustrated in  Fig.~\ref{non-crossing}, where qubit $q_2$ is relocated to an SLM trap to avoid AOD crossing. This transition from an AOD trap to an SLM trap can be achieved by reducing the intensity of the AOD trap while either enhancing or maintaining the intensity of the SLM trap. After this operation, $q_2$ in the SLM trap no longer requires vertical or horizontal acousto-optic deflectors to maintain its position. Next, within the AOD array, $c_0$ is shifted rightward, and $r_0$ is moved downward, allowing $q_0$ to be transported near $q_1$ at a distance smaller than the Rydberg blockade radius $r_b$. At this point, $q_0$ and $q_1$ can form an entanglement gate.

Notably, as long as the row/column non-crossing constraint of the AOD array is maintained, different sections of the array can operate independently. This characteristic enables neutral atom quantum computing with significant flexibility and reconfigurability. In addition to the AOD crossing constraint, neutral atom quantum computers must also adhere to the parallel execution constraint, which dictates that when two qubits undergo a two-qubit gate operation, no third qubit can be present within a certain vicinity determined by the Rydberg blockade radius $r_b$ \cite{Tan2022, Li2023}.

\subsection{Related work}

To effectively address these constraints and establish a well-defined compilation process for reconfigurable neutral atom arrays, several methods have been proposed. Tan et al. \cite{Tan2022} first discretized the working space of the quantum compilation problem by partitioning space into an array of sites that are sufficiently spaced apart and restricting quantum gate execution to operate in a single site. This approach effectively mitigates issues arising from the restriction zone. Based on this architecture, they also introduced the first solver-based compilation method for reconfigurable neutral atom arrays. Building upon this work, Tan et al. \cite{Tan2024} proposed DPQA, a compilation method that integrates a greedy search algorithm with solver-based approaches to improve both computational efficiency and scalability. This method enhances the scalability of the original solver-based approach, further optimizing the compilation of reconfigurable neutral atom quantum circuits.

\begin{table}[t]
	\renewcommand{\arraystretch}{1}
	\caption{Compilation time of DPQA.}
	\label{tab:dpqa-time}
	\footnotesize
	\centering
	\tabcolsep=0.02\linewidth
	\begin{tabular}{ccccc}
		\toprule
		Quantum Circuit & Qubits & Gates & Depth & Compilation Time(s) \\ 
		\midrule
		rand3reg\_70\_0 & 70 & 105 & 15 &2906.7 \\
		rand3reg\_70\_1 & 70 & 105 & 12 &3195.9 \\
		rand3reg\_70\_2 & 70 & 105 & 14 &3573.4 \\
		rand3reg\_70\_3 & 70 & 105 & 14 &3312.3 \\
		rand3reg\_70\_4 & 70 & 105 & 15 &2848.3 \\
		rand3reg\_80\_3 & 70 & 105 & 12 &9696.3 \\
		rand3reg\_80\_4 & 70 & 105 & 14 &8088.4 \\
		rand3reg\_80\_7 & 70 & 105 & 16 &8756.2 \\
		rand3reg\_80\_8 & 70 & 105 & 18 &9334.1 \\
		rand3reg\_80\_9 & 70 & 105 & 15 &9630.4 \\
		\bottomrule
	\end{tabular}
\end{table}

Although DPQA significantly accelerates the compilation process for neutral atom quantum computers, it still suffers from severe computational efficiency issues. The compilation times for $10$ circuits containing $70$ to $80$ qubits, as shown in Table~\ref{tab:dpqa-time}, range from approximately $2848$ seconds to over $9600$ seconds. Such excessive compilation times lead to inefficient utilization of quantum computing resources, which is an increasingly pressing concern as qquantum circuits continue to grow in size. Furthermore, as neutral atom quantum computer technology advances, the solution space for neutral quantum compilation becomes dramatically expansive with the increasing scale of 2D arrays, and current neutral atom compilation techniques still exhibit significant limitations in addressing this challenge.

\section{Physics-Aware Compilation for Parallel Quantum Circuit Execution} \label{sec:pac-method}

\subsection{Key issues and methods overviews}

In the field of neutral atom quantum compilation, various methods have been proposed to address hardware-specific constraints. Among these approaches, solver-based methods have demonstrated significantly better results compared to heuristic approaches, making them widely adopted in practical applications. However, since the solver is based on the Max-SAT problem which is NP-Hard \cite{Bonet2007}, solver efficiency represents a critical challenge in these methods. As shown in Table~\ref{tab:dpqa-time}, compilation times for moderately complex circuits can reach thousands of seconds, imposing prohibitive computational costs that become increasingly severe with larger system sizes. Combined with the flexibility of neutral atom hardware, this leads to a technical tension where maintaining the inherent reconfigurability of these systems while achieving acceptable compilation performance becomes challenging. To fully leverage the advantages of neutral atom quantum computers, more efficient approaches are needed that preserve hardware flexibility while delivering practical compilation times.

We also observe that applying solvers across the entire computational plane creates unnecessary coupling between regions that could be processed independently. This superfluous coupling reduces compilation efficiency by preventing naturally separable quantum operations from being executed in parallel, despite the inherent physical independence provided by the quantum hardware architecture.

Our analysis suggests an alternative approach that integrates hardware physics understanding into the compilation process. Rather than treating compilation as purely computational, we recognize that the physical characteristics of neutral atom hardware provide natural opportunities for problem decomposition. The physical constraints of neutral atom arrays create natural boundaries that can be leveraged to partition both the hardware plane and the quantum circuit, reducing the solution space without sacrificing hardware flexibility.

\begin{figure}[h]
	\centering
	\includegraphics[width=1\textwidth, trim=1.8cm 7.5cm 1.8cm 0cm, clip]{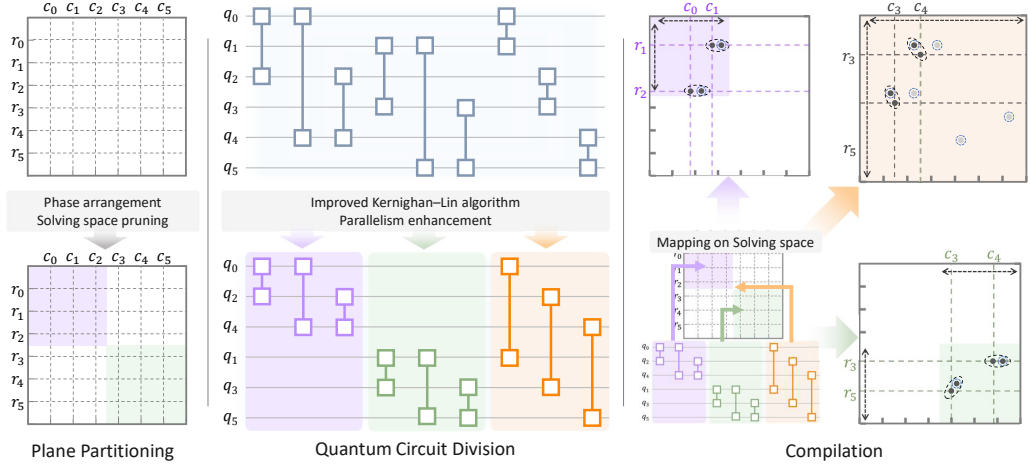}
	\caption{Workflow of PAC.}
	\label{fig:main}
\end{figure}

Based on these insights, we propose  Physics-Aware Compilatio (PAC), a method that addresses these challenges by integrating hardware physics into the compilation process. As shown in Fig.~\ref{fig:main}, PAC consists of two sequential components:

\begin{itemize}
	\item \textbf{Hardware Plane Partitioning} divides the 2D hardware plane into independent sections based on physical device characteristics. This partitioning leverages the natural boundaries created by physical constraints to reduce the solution space while enabling parallelism in the compilation process.
	\item \textbf{Quantum Circuit Division} transforms the quantum circuit compilation problem through an improved graph-based partitioning approach. By dividing the circuit into communities with stronger internal connections, this component enables more efficient compilation by aligning with the underlying physical structure of quantum operations.
\end{itemize}

These techniques work together to improve compilation efficiency while preserving the hardware flexibility that makes neutral atom platforms more attractive for quantum computing applications.

\subsection{Hardware Plane Partitioning}
\label{subsec:HPP}

\begin{figure}[h]
	\centering
	\includegraphics[width=1\textwidth, trim=0cm 2.5cm 4cm 1.5cm, clip]{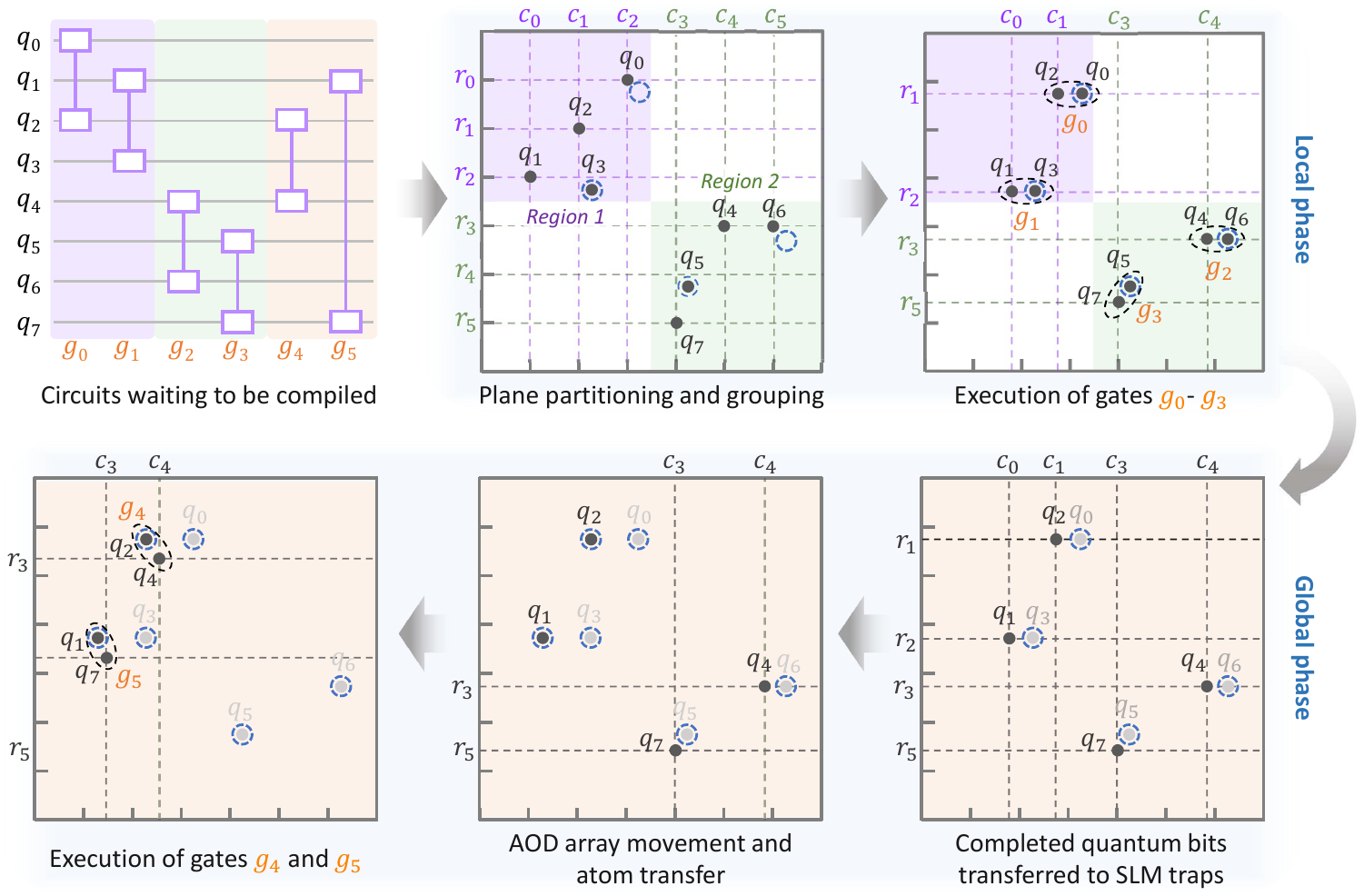}
	\caption{Demonstration of hardware plane partitioning.}
	\label{fig:hpp}
\end{figure}

Hardware plane partitioning is demonstrated using a $6\times 6$ neutral atom array in Fig.\ref{fig:hpp}, addressing efficiency challenges through strategic resource division. The process comprises two sequential phases: local and global, operating on the target quantum circuit.

In the local phase, we partition the hardware plane into two equal-sized regions, as shown in Figure \ref{fig:hpp}: Region 1 (purple, upper left) and Region 2 (green, lower right). The AOD array is similarly divided into two groups: the first includes columns ${c_0,c_1,c_2}$ and rows ${r_0,r_1,r_2}$, while the second includes columns ${c_3,c_4,c_5}$ and rows ${r_3,r_4,r_5}$. This partitioning leverages a key physical property of the hardware: AOD arrays in separate regions can operate independently without interference, enabling true parallelism in the compilation process. During this phase, we perform initial mapping by dividing the logical qubits between the two regions and then compile all quantum operations that can be executed entirely within either region like $g_0-g_3$ .

During compilation, we classify qubits in each region into two types based on their interaction in quantum circuit: resolved qubits and active qubits. Resolved qubits (such as $q_0$, $q_3$, $q_5$ and $q_6$ in Figure \ref{fig:hpp}) can complete all their required operations entirely within their assigned region during the local phase. Active qubits ($q_1$, $q_2$, $q_4$ and $q_7$) require interactions with qubits in other regions and therefore need additional operations in the global phase to complete their quantum gates.

Global phase compilation requires all resolved qubits to be placed in SLM traps by the local phase conclusion, as shown in Figure \ref{fig:hpp} where resolved qubits appear in gray. This arrangement is critical: SLM traps do not occupy AOD grid positions, enabling active qubits to move freely during the global phase without obstruction. This requirement is implemented through appropriate constraints in the local phase compilation. The global phase inherits qubit positions and states from the local phase as initial conditions, then completes compilation by processing remaining operations across the entire plane. By focusing exclusively on active qubits and their cross-region interactions, this approach reduces problem complexity compared to traditional methods that process all qubits simultaneously. The final compilation solution merges both phases' results, with total circuit depth calculated as the maximum depth between the two local phases plus the depth of the global phase. Algorithm \ref{alg:hpp} presents the complete hardware plane partitioning process as follow.

\begin{algorithm}[H]
	\footnotesize
	\setstretch{1.2}
	\caption{Hardware Plane Partitioning} 
	\label{alg:hpp}
	\KwIn{A $N\times N$ Neutral Atom Array $A$, A Commutable Quantum Circuit $QC=\{Q,E\}$} 
	\KwOut{Compilation result $R$} 
	// Partition the hardware plane into two sub-arrays\\
	$A_1 = A[0:\lceil \frac{N}{2} \rceil-1, 0:\lceil \frac{N}{2} \rceil-1]$;\\
	$A_2 = A[\lceil \frac{N}{2} \rceil:N,\lceil \frac{N}{2} \rceil:N]$;\\
	// Decompose QC into three subcircuits:\\
	$QC_1=(Q_1, E_1), QC_2=(Q_2, E_2), QC_3=(Q_3, E_3), Q_{a1}, Q_{a2} \leftarrow$ Partition(QC) \\
	where $Q_{a1} = Q_1 \setminus Q(E_3)$, $Q_{a2} = Q_2 \setminus Q(E_3)$;\\
	// Local Phase Compilation\\
	$R_1 \leftarrow$ Compile($QC_1, A_1$), with all $Q_{a1}$ qubits placed in SLM traps in the last layer;\\
	$R_2 \leftarrow$ Compile($QC_2, A_2$), with all $Q_{a2}$ qubits placed in SLM traps in the last layer;\\
	// Global Phase Compilation \\
	\For{$q\in Q$}{
		Retrieve $q$’s state and position from $A$;\\
		Update $R_1$ and $R_2$ accordingly;\\
	}
	$R_3 \leftarrow$ Compile($QC_3, A$), inheriting qubit states and positions from $R_1$ and $R_2$;\\
	// Merge results\\
	$R \leftarrow Merge(R_1, R_2, R_3)$;
	Return R;
\end{algorithm}

The efficiency advantage of hardware plane partitioning stems from two intuitive but effective key factors. First, hardware-level partitioning significantly prunes the solution space by reducing the number of variables and constraints that must be handled by the solver in each phase. Instead of solving one large optimization problem across the entire hardware plane, we solve multiple smaller problems that are computationally more tractable. Second, the independence of regions in the local phase enables parallel compilation, further reducing the overall time required.

\subsection{Quantum Circuit Division} \label{subsec:QCD}

Quantum circuit division addresses the second key component of our physics-aware approach, focusing on how to effectively partition quantum operations to maximize parallel execution capabilities. The goal is to determine which gates should be executed in the local phase versus the global phase to minimize overall compilation time, defined as $T = T_1+ T_2$, where $T_1$ and $T_2$ correspond to the compilation time for the local and global phases, respectively. With $N_G$ total quantum gates requiring compilation, this becomes an allocation problem of distributing gates between phases. Since the hardware plane partitioning enables parallel execution in the local phase with reduced search space, strategies for component prioritizes maximizing the number of gates assigned to the local phase while minimizing cross-region interactions that must be handled in the global phase.

\begin{figure*}[t]
	\centering
	\includegraphics[width=1\textwidth, trim=3cm 13.5cm 2.2cm 0cm, clip]{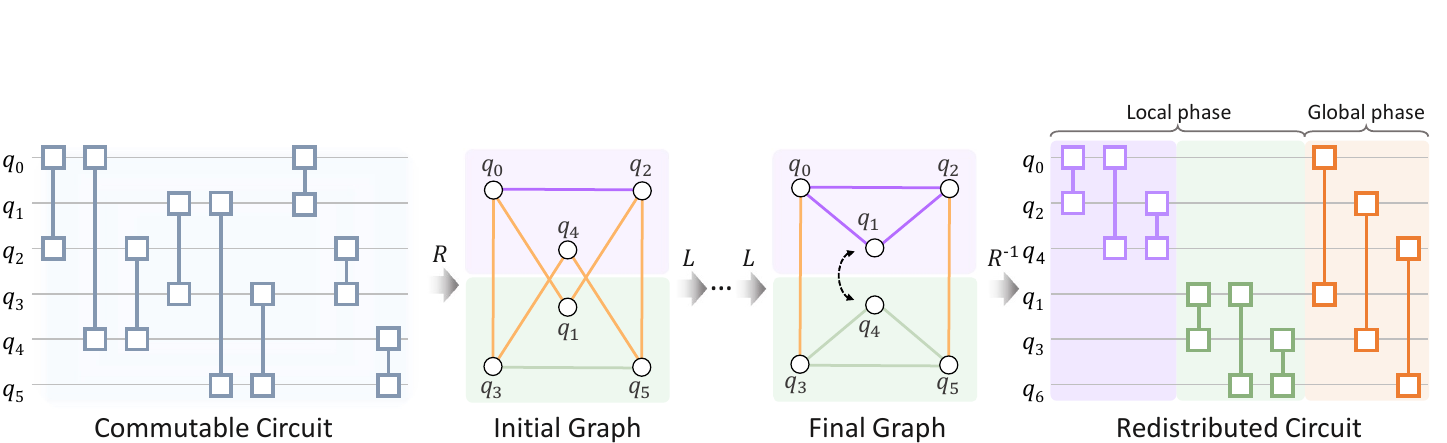}
	\caption{Demonstration of quantum circuit division.}
	\label{fig:qcd}
\end{figure*}

To achieve effective quantum circuit division, we propose an improved Kernighan–Lin algorithm that aligns with the physical structure of the neutral atom hardware. As illustrated in Fig. \ref{fig:qcd}, algorithm begins by representing the quantum circuit as a graph where each qubit corresponds to a vertex and each two-qubit gate corresponds to an edge. This graph representation naturally captures the physical interaction patterns between qubits. The vertices are then evenly partitioned into two communities, corresponding to the two regions of the hardware plane. Within each community, qubits are classified into resolved qubits and active qubits as previously defined in Section \ref{subsec:HPP}.  The process determine resolved qubits and active qubits is shown in Algorithm \ref{alg:qcd_1} as follow.

\begin{algorithm}[H]
	\footnotesize
	\setstretch{1.25}
	\caption{Identify initial resolved qubits and active qubits} 
	\label{alg:qcd_1}
	\KwIn{A Commutable Quantum Circuit $QC=\{Q,E\}$, Weight parameter $k$, Maximum number of swaps $Max\_iter$} 
	\KwOut{Quantum circuit partitions for local and global phase compilation $QC_1$, $QC_2$, $QC_3$} 
	$Q_1$ = RandomSubset($Q$, $\lceil|Q|/2\rceil$);\\
	$Q_2$ = $Q\backslash Q_1$;\\
	$swap\_flag$ = True;\\
	$n\_iter$ = 0;\\
	$E_1$, $E_2$, $E_3$ = \{\}, \{\}, \{\};\\
	Active qubits $Q_{a1}$, $Q_{a2}$ = \{\}, \{\};\\
	\For{e in $E$}{
		\uIf{e[0] in $Q_1$ and e[1] in $Q_1$}{
			$E_1$ = $E_1\cup \{e\}$;\\
		}
		\uElseIf{e[0] in $Q_2$ and e[1] in $Q_2$}{
			$E_2$ = $E_2\cup \{e\}$;\\
		}
		\uElseIf{e[0] in $Q_1$ and e[1] in $Q_2$}{
			$E_3$ = $E_3\cup \{e\}$;\\
		}
		\uElseIf{e[0] in $Q_1$ and e[1] in $Q_2$}{
			$E_3$ = $E_3\cup \{e\}$;\\
			$Q_{a1}\cup \{e[0]\}$;\\u
			$Q_{a2}\cup \{e[1]\}$;\\
		}
		\ElseIf{e[0] in $Q_2$ and e[1] in $Q_1$}{
			$E_3$ = $E_3\cup \{e\}$;\\
			$Q_{a1}\cup \{e[1]\}$;\\
			$Q_{a2}\cup \{e[0]\}$;\\
		}
	}
	$Q_{r1}=Q_1\backslash Q_{a1}$;\\
	$Q_{r2}=Q_2\backslash Q_{a2}$;\\
	return Resolved qubits $Q_{r1}$, $Q_{r2}$, Active qubits $Q_{a1}$, $Q_{a2}$, Edge sets $E_1$, $E_2$, $E_3$
\end{algorithm}

Next, candidates for community reassignment are identified by examining active qubits' connectivity patterns. Qubits with more external connections than internal ones are defined as swap qubits, indicating suboptimal placement from a physical execution perspective, since they interact more with qubits outside their region than within it. If one community has no swap qubits while another does, active qubits from the empty-set community are designated as swap qubits to enable balanced optimization.

We then iteratively optimize the community structure by swapping pairs of swap qubits between communities, computing the loss function $L$ for each potential swap. This process directly improves the physical execution efficiency by minimizing cross-region interactions that would require more complex global phase compilation. The pair of qubits that minimizes $L$ is selected for swapping, followed by an update of resolved and active qubit classifications. This optimization process continues until one of three termination conditions is met: (a) both communities have empty swap qubit sets, indicating an optimal or near-optimal partition has been achieved; (b) the number of swaps reaches a predefined upper limit to prevent excessive computation; (c) no swap operation can further reduce the loss function $L$, suggesting we've reached a local optimum in the partition quality.

The loss function $L$ is defined as:

\begin{equation}
	\label{equ:L}
	L = k(N_{f,1}+N_{f,2})+(1-k)(N_{E,cross})
\end{equation}

Here, $N_{f1}$ and $N_{f2}$ represent the number of active qubits in each of the two communities that require cross-region interactions, while $N_{E,cross}$ denotes the number of edges (two-qubit gates) between these communities. The parameter $k\in [0,1]$ balances the optimization priority between minimizing active qubits and minimizing cross-region gates. This physics-aware loss function directly targets the efficiency bottleneck: by minimizing $L$, we reduce both the number of qubits and the number of gates that must be processed in the computationally expensive global phase, thereby maximizing the operations that can be executed in parallel during the local phase.

After optimization, we convert the resulting graph back into a quantum circuit and mark each gate for execution in either the local or global phase based on the final community structure. This physics-guided partitioning transforms the conventional global compilation problem into a more efficient region-based approach. The improved Kernighan–Lin algorithm is shown in Algorithm \ref{alg:qcd}.

\begin{algorithm}[H]
	\footnotesize
	\setstretch{1}
	\caption{The improved Kernighan–Lin algorithm} 
	\label{alg:qcd}
	\KwIn{A Commutable Quantum Circuit $QC=\{Q,E\}$, Resolved qubits $Q_{r1}$, $Q_{r2}$, Active qubits $Q_{a1}$, $Q_{a2}$, Edge sets $E_1$, $E_2$, $E_3$, Weight parameter $k$, Maximum number of swaps $Max\_iter$} 
	\KwOut{Quantum circuit partitions $QC_1$, $QC_2$, $QC_3$} 
	$swap\_flag$ = True;\\
	$n\_iter$ = 0;\\
	$swap\_flag$ = False;\\
	\While{True}{
		// Identify swap qubits\\
		$Q_{s1}, Q_{s2}$=\{\}, \{\};\\
		\For{q in $Q_{a1}$}{
			\If{$|\{e \in E_3|q \in e\}|\leq |\{e \in E_1| q \in e\}| $}{
				$Q_{s1} \cup \{q\}$;\\
			}
		}
		\For{q in $Q_{a2}$}{
			\If{$|\{e \in E_3|q \in e\}|\leq |\{e \in E_2| q \in e\}| $}{
				$Q_{s2} \cup \{q\}$;\\
			}
		}
		\If{$Q_{s1}$ == \{\}}{$Q_{s1}=Q_{a1}$;}
		\If{$Q_{s2}$ == \{\}}{$Q_{s2}=Q_{a2}$;}
		$min\_L$=$k(|Q_{a1}|+|Q_{a2}|) + (1-k)|E_3|$;\\
		$best\_q1$, $best\_q2$ = None, None;\\
		\For{each $(q_1,q_2) \in Q_{s1} \times Q_{s2}$}{
			Try swap $q_1$, $q_2$ $\Rightarrow$ and compute $L$;\\
			\If{$L\leq min\_L$}{
				$best\_q1$, $best\_q2$ = $q_1$, $q_2$;\\
				$min\_L$ = $L$;\\
			}
		}
		\If{$best\_q1$ is not None and $best\_q2$ is not None}{
			Swap $best\_q1$ and $best\_q2$;\\
			$swap\_flag$ = True;\\
			$n\_iter$ = $n\_iter$ + 1;\\
		}
		\If{($Q_{s1}$ == \{\} and $Q_{s2}$ == \{\}) or not $swap\_flag$ or $n\_iter \leq Max\_iter$}{
			break;}
	}
	return $QC_1=(Q_1,E_1)$, $QC_2=(Q_2,E_2)$, $QC_3=(Q_3,E_3)$, $Q_{a1}$, $Q_{a2}$, $Q_{r1}$, $Q_{r2}$
\end{algorithm}

At this stage, the quantum circuit is partitioned into three segments: $QC_1$, $QC_2$, and $QC_3$. The qubits involved in $QC_1$ and $QC_2$ are categorized into active qubits ($Qa_1$, $Qa_2$) and resolved qubits ($Qr_1$, $Qr_2$). According to the hardware plane partitioning strategy, $QC_1$ and $QC_2$ are executed in the local phase, while $QC_3$ is assigned to the global phase. As described in Section~\ref{subsec:HPP}, the local phase serves as the initial stage of solving, where $QC_1$ and $QC_2$ are processed in parallel.

To prevent interference from resolved qubits in the local phase during the subsequent global phase, we impose specific constraints during local-phase solving. To clearly illustrate our method, we adopt the same variable definitions as DPQA\cite{Tan2024}. Specifically, we define $x_{i,t}$ and $y_{i,t}$ as the SLM coordinates of qubit $i$ at time $t$, while $c_{i,t}$ and $r_{i,t}$ denote the corresponding AOD column and row indices. The binary variable $a_{i,t}$ indicates the state of qubit $i$ at time $t$, where $a=0$ represents the SLM state and $a=1$ represents the AOD state.

In the local phase, if all remaining gates are resolved at a given solving step, we impose the following constraint on all resolved qubits to avoid unnecessary interaction in the global phase:

\begin{equation}
	\forall i \in Q_{r1},\ \forall t,\ \text{if } t \text{ is the final stage},\ a_{i,t} = 0
\end{equation}

The above constraint effectively prevents resolved qubits from interfering with the global phase and reduces the number of variables involved in global-phase solving.

In the global phase, $QC_3$ is solved over the entire neutral atom array. To ensure continuity and reduce the solution space, we introduce two additional constraints. First, since each SLM trap can accommodate at most one atom, no qubit in the global phase is allowed to occupy an SLM position already assigned to a qubit in the local phase:

\begin{equation}
	\forall t,\ \forall (x_s, y_s) \in \mathcal{F},\quad \neg(x_{i,t} = x_s \land y_{i,t} = y_s \land a_{i,t} = 0)
\end{equation}

where $\mathcal{F}$ denotes the set of SLM coordinates of $Q_{r1}$ and $Q_{r2}$ in the local phase. To ensure consistency between the local phase and the global phase, the initial configuration of the global phase must satisfy the final state of the local phase, including both SLM and AOD assignments.
The SLM coordinates of each qubit in the global phase must match its final position in the local phase. Let $(x_q^L, y_q^L)$ denote the final SLM coordinates of qubit $q$ in active qubits of local phase. Then:

\begin{equation} 
	x_{q,0} = x_q^L \land y_{q,0} = y_q^L,\quad \forall q \in Q_{a1} \cup Q_{a2}
\end{equation}

The AOD channel assignment in the global phase must preserve the relative ordering of qubits established in the local phase. Let $c_q^L$ represent the final AOD index of qubit $q$ in the local phase, and $c_{q,0}$ its AOD index at time $t = 0$ in the global phase. Then, for any pair of qubits $q_1, q_2 \in Q_{a1} \cup Q_{a2}$, the following conditions must hold:

\begin{equation}
	\left\{
	\begin{aligned}
		c_{q_1,0} < c_{q_2,0}, & \quad \text{if } c_{q_1}^L < c_{q_2}^L \\
		c_{q_1,0} > c_{q_2,0}, & \quad \text{if } c_{q_1}^L > c_{q_2}^L \\
		c_{q_1,0} = c_{q_2,0}, & \quad \text{if } c_{q_1}^L = c_{q_2}^L
	\end{aligned}
	\right.
	\quad \text{for all } q_1, q_2 \in Q_{a1} \cap Q_{a2}
\end{equation}

These constraints are essential to guarantee the correctness of the compilation process and to ensure seamless continuity between the local and global phases.

\section{Experimental Results}
\subsection{Experimental Setup}

PAC was evaluated against DPQA \cite{Tan2024}, the current state-of-the-art method for neutral atom quantum compilation. DPQA has demonstrated performance improvements over previous approaches such as RAA \cite{Tan2022}, making it an appropriate baseline for comparison. The evaluation utilized $40$ random $3$-regular graph circuits with $60-90$ qubits and $90-135$ two-qubit gates, identical to those used in DPQA's evaluation. These circuits are applicable to various combinatorial optimization problems including Max-Cut and quantum simulation algorithms such as QAOA and trotterized adiabatic evolution \cite{Farhi2000, Ebadi2022}. To assess scalability, performance was tested across multiple neutral atom array dimensions: $16\times16$, $20\times20$, $24\times24$, $28\times28$, $50\times50$, and $64\times64$. The $16\times16$ configuration represents currently proposed experimental systems, while larger dimensions reflect anticipated future hardware capabilities. Performance was measured using two key metrics: compilation efficiency through solution time and speedup ratio compared to baseline with a $10,000$-second timeout threshold, and compilation quality through quantum circuit layer count and layer reduction ratio defined as follow:

\begin{equation}
	\label{equ:layer-reduction-ratio}
	R_{cl} = \frac{NL_{PAC}-NL_{baseline}}{NL_{baseline}}
\end{equation}

where $NL_{PAC}$ and $NL_{baseline}$ are number of circuit layers after PAC and baseline, respectively. Detailed information about the experimental setup is provided in Appendix B.

\subsection{Solving Efficiency and Quality}

Table \ref{tab:16x16} presents the compilation performance comparison between PAC and DPQA on a $16\times16$ AOD array. PAC demonstrates substantial efficiency improvements, compiling all test circuits within $1200$ seconds, while DPQA encounters limitations when dealing with circuits beyond $70$ qubits. For circuits where both methods complete, PAC shows a notable speedup of up to $78.5\times$ with an average acceleration of $53.5\times$, corresponding to a significant reduction in compilation time. Importantly, this performance gain does not compromise compilation quality, as PAC maintains an average circuit depth of $12.6$, comparable to DPQA's results across most circuits.

The performance advantage becomes more pronounced on larger architectures, as shown in Table \ref{tab:64x64} for a $64\times64$ array. On this larger configuration, PAC achieves even greater efficiency with a peak speeds up to $139.16\times$ and completes all test circuits in under $200$ seconds. The average speedup across DPQA-solvable circuits increases to $78.23\times$. Despite this substantial acceleration, PAC maintains high-quality results with an average circuit depth of $12.76$ compared to DPQA's $12.8$, demonstrating that the performance improvements do not come at the expense of compilation quality.

A clear trend emerges when examining PAC's performance across different array sizes: compilation efficiency improves as hardware scale increases. While maximum compilation time reaches approximately $1200$ seconds on $16\times16$ arrays, it decreases to under $200$ seconds on $64\times64$ arrays, indicating excellent scalability with hardware size. This positive scaling relationship suggests that PAC's advantage becomes more valuable as quantum hardware continues to advance toward larger scales. Results for intermediate array sizes ($20\times20$, $24\times24$, $28\times28$, and $50\times50$) follow the same trend and are presented in \ref{Experimentation on series array size}.

\newpage

\begin{table}[H]
	\footnotesize
	\caption{Solving Time and Circuit Layers on 16$\times$16 Array.}
	\label{tab:16x16}
	\centering
	\tabcolsep=0.01\linewidth
	\renewcommand{\arraystretch}{1.}
	\begin{tabular}{ccccccccc}
		\toprule
		\multirow{2}{*}{Quantum Circuit} & \multirow{2}{*}{Qubits} & \multicolumn{3}{c}{Solving time (s)}  & \multicolumn{4}{c}{Circuit Layers} \\
		\cmidrule(lr){3-5}\cmidrule(lr){6-9}
		&                         & DPQA & PAC      & Speedup & DPQA & PAC  & $\Delta _{cl}$   & $R_{cl}$      \\
		\midrule
		rand3reg\_60\_0 & 60 & 1728.42 & 30.40   & \textbf{56.86 $\uparrow$}                 & 12 & 13   & 1   & 8.33\%  \\
		rand3reg\_60\_1 & 60 & 2701.22 & 39.99   & \textbf{67.54 $\uparrow$}                 & 11 & 12   & 1   & 9.09\%  \\
		rand3reg\_60\_2 & 60 & 1281.69 & 50.71   & \textbf{25.27 $\uparrow$}                 & 10 & 11   & 1   & 10.00\% \\
		rand3reg\_60\_3 & 60 & 1968.34 & 35.16   & \textbf{55.98 $\uparrow$}                 & 10 & 11   & 1   & 10.00\% \\
		rand3reg\_60\_4 & 60 & 1817.24 & 30.81   & \textbf{58.98 $\uparrow$}                 & 12 & 12   & 0   & 0.00\%  \\
		rand3reg\_60\_5 & 60 & 1420.21 & 48.28   & \textbf{29.41 $\uparrow$}                 & 12 & 12   & 0   & 0.00\%  \\
		rand3reg\_60\_6 & 60 & 1645.56 & 48.93   & \textbf{33.63 $\uparrow$}                 & 12 & 12   & 0   & 0.00\%  \\
		rand3reg\_60\_7 & 60 & 1636.04 & 37.72   & \textbf{43.38 $\uparrow$}                 & 10 & 11   & 1   & 10.00\% \\
		rand3reg\_60\_8 & 60 & 1871.13 & 35.61   & \textbf{52.55 $\uparrow$}                 & 11 & 12   & 1   & 9.09\%  \\
		rand3reg\_60\_9 & 60 & 1853.53 & 40.75   & \textbf{45.49 $\uparrow$}                 & 13 & 13   & 0   & 0.00\%  \\
		rand3reg\_70\_0 & 70 & 5539.06 & 75.14   & \textbf{73.72 $\uparrow$}                 & 13 & 13   & 0   & 0.00\%  \\
		rand3reg\_70\_1 & 70 & 5746.01 & 139.23  & \textbf{41.27 $\uparrow$}                 & 12 & 12   & 0   & 0.00\%  \\
		rand3reg\_70\_2 & 70 & 4353.12 & 82.84   & \textbf{52.55 $\uparrow$}                 & 13 & 13   & 0   & 0.00\%  \\
		rand3reg\_70\_3 & 70 & 6270.31 & 97.65   & \textbf{64.21 $\uparrow$}                 & 13 & 14   & 1   & 7.69\%  \\
		rand3reg\_70\_4 & 70 & 6873.92 & 117.21  & \textbf{58.64 $\uparrow$}                 & 12 & 12   & 0   & 0.00\%  \\
		rand3reg\_70\_5 & 70 & 5999.32 & 76.40   & \textbf{78.52 $\uparrow$}                 & 14 & 14   & 0   & 0.00\%  \\
		rand3reg\_70\_6 & 70 & 6334.61 & 87.66   & \textbf{72.27 $\uparrow$}                 & 13 & 14   & 1   & 7.69\%  \\
		rand3reg\_70\_7 & 70 & 6214.34 & 137.41  & \textbf{45.22 $\uparrow$}                 & 12 & 14   & 2   & 16.67\% \\
		rand3reg\_70\_8 & 70 & 6179.55 & 97.32   & \textbf{63.50 $\uparrow$}                 & 14 & 14   & 0   & 0.00\%  \\
		rand3reg\_70\_9 & 70 & 3847.52 & 75.23   & \textbf{51.15 $\uparrow$}                 & 11 & 13   & 2   & 18.18\% \\
		rand3reg\_80\_0 & 80 & TO      & 559.48  & \textbf{$\textgreater{}$17.87 $\uparrow$} & -  & 15   & -   & -       \\
		rand3reg\_80\_1 & 80 & TO      & 239.20  & \textbf{$\textgreater{}$41.81 $\uparrow$} & -  & 15   & -   & -       \\
		rand3reg\_80\_2 & 80 & TO      & 360.70  & \textbf{$\textgreater{}$27.72 $\uparrow$} & -  & 15   & -   & -       \\
		rand3reg\_80\_3 & 80 & TO      & 964.44  & \textbf{$\textgreater{}$10.37 $\uparrow$} & -  & 15   & -   & -       \\
		rand3reg\_80\_4 & 80 & TO      & 277.58  & \textbf{$\textgreater{}$36.03 $\uparrow$} & -  & 14   & -   & -       \\
		rand3reg\_80\_5 & 80 & TO      & 769.42  & \textbf{$\textgreater{}$13.00 $\uparrow$} & -  & 15   & -   & -       \\
		rand3reg\_80\_6 & 80 & TO      & 287.65  & \textbf{$\textgreater{}$34.76 $\uparrow$} & -  & 15   & -   & -       \\
		rand3reg\_80\_7 & 80 & TO      & 152.97  & \textbf{$\textgreater{}$65.37 $\uparrow$} & -  & 14   & -   & -       \\
		rand3reg\_80\_8 & 80 & TO      & 545.00  & \textbf{$\textgreater{}$18.35 $\uparrow$} & -  & 15   & -   & -       \\
		rand3reg\_80\_9 & 80 & TO      & 341.63  & \textbf{$\textgreater{}$29.27 $\uparrow$} & -  & 15   & -   & -       \\
		rand3reg\_90\_0 & 90 & TO      & 674.18  & \textbf{$\textgreater{}$14.83 $\uparrow$} & -  & 16   & -   & -       \\
		rand3reg\_90\_1 & 90 & TO      & 561.58  & \textbf{$\textgreater{}$17.81 $\uparrow$} & -  & 16   & -   & -       \\
		rand3reg\_90\_2 & 90 & TO      & 784.85  & \textbf{$\textgreater{}$12.74 $\uparrow$} & -  & 18   & -   & -       \\
		rand3reg\_90\_3 & 90 & TO      & 668.09  & \textbf{$\textgreater{}$14.97 $\uparrow$} & -  & 18   & -   & -       \\
		rand3reg\_90\_4 & 90 & TO      & 583.63  & \textbf{$\textgreater{}$17.13 $\uparrow$} & -  & 18   & -   & -       \\
		rand3reg\_90\_5 & 90 & TO      & 642.10  & \textbf{$\textgreater{}$15.57 $\uparrow$} & -  & 16   & -   & -       \\
		rand3reg\_90\_6 & 90 & TO      & 1164.81 & \textbf{$\textgreater{}$8.59 $\uparrow$}  & -  & 20   & -   & -       \\
		rand3reg\_90\_7 & 90 & TO      & 455.63  & \textbf{$\textgreater{}$21.95 $\uparrow$} & -  & 18   & -   & -       \\
		rand3reg\_90\_8 & 90 & TO      & 671.65  & \textbf{$\textgreater{}$14.89 $\uparrow$} & -  & 18   & -   & -       \\
		rand3reg\_90\_9 & 90 & TO      & 903.11  & \textbf{$\textgreater{}$11.07 $\uparrow$} & -  & 18   & -   & -       \\
		&    & 3764.06 & 69.22   & \textbf{53.51 $\uparrow$}                 & 12 & 12.6 & 0.6 & 5.00\%   \\
		\bottomrule
	\end{tabular}
\end{table}

\newpage

\begin{table}[H]
	\footnotesize
	\caption{Solving Time and Circuit Layers on 64$\times$64 Array.}
	\label{tab:64x64}
	\centering
	\tabcolsep=0.01\linewidth
	\renewcommand{\arraystretch}{1.0}
	\begin{tabular}{ccccccccc}
		\toprule
		\multirow{2}{*}{Quantum Circuit} & \multirow{2}{*}{Qubits} & \multicolumn{3}{c}{Solving time (s)}  & \multicolumn{4}{c}{Circuit Layers} \\
		\cmidrule(lr){3-5}\cmidrule(lr){6-9}
		&                         & DPQA & PAC      & Speedup & DPQA & PAC  & $\Delta _{cl}$   & $R_{cl}$      \\
		\midrule
		rand3reg\_60\_0 & 60 & 1316.52 & 18.19  & \textbf{72.37 $\uparrow$}                  & 12   & 11    & \textbf{-1}    & \textbf{-8.33\%} \\
		rand3reg\_60\_1 & 60 & 1040.71 & 23.62  & \textbf{44.06 $\uparrow$}                  & 12   & 12    & 0              & 0.00\%           \\
		rand3reg\_60\_2 & 60 & 1411.00 & 22.56  & \textbf{62.55 $\uparrow$}                  & 11   & 11    & 0              & 0.00\%           \\
		rand3reg\_60\_3 & 60 & 1528.79 & 24.22  & \textbf{63.13 $\uparrow$}                  & 12   & 12    & 0              & 0.00\%           \\
		rand3reg\_60\_4 & 60 & 892.77  & 18.01  & \textbf{49.56 $\uparrow$}                  & 11   & 10    & \textbf{-1}    & \textbf{-9.09\%} \\
		rand3reg\_60\_5 & 60 & 1525.45 & 22.12  & \textbf{68.96 $\uparrow$}                  & 12   & 12    & 0              & 0.00\%           \\
		rand3reg\_60\_6 & 60 & 953.52  & 20.08  & \textbf{47.47 $\uparrow$}                  & 12   & 12    & 0              & 0.00\%           \\
		rand3reg\_60\_7 & 60 & 1337.51 & 21.76  & \textbf{61.45 $\uparrow$}                  & 11   & 12    & 1              & 9.09\%           \\
		rand3reg\_60\_8 & 60 & 1070.16 & 19.12  & \textbf{55.98 $\uparrow$}                  & 11   & 11    & 0              & 0.00\%           \\
		rand3reg\_60\_9 & 60 & 873.20  & 22.11  & \textbf{39.49 $\uparrow$}                  & 12   & 12    & 0              & 0.00\%           \\
		rand3reg\_70\_0 & 70 & 2906.76 & 41.35  & \textbf{70.29 $\uparrow$}                  & 13   & 13    & 0              & 0.00\%           \\
		rand3reg\_70\_1 & 70 & 3195.99 & 47.87  & \textbf{66.76 $\uparrow$}                  & 13   & 13    & 0              & 0.00\%           \\
		rand3reg\_70\_2 & 70 & 3573.35 & 43.42  & \textbf{82.31 $\uparrow$}                  & 13   & 13    & 0              & 0.00\%           \\
		rand3reg\_70\_3 & 70 & 3312.31 & 36.53  & \textbf{90.68 $\uparrow$}                  & 14   & 13    & \textbf{-1}    & \textbf{-7.14\%} \\
		rand3reg\_70\_4 & 70 & 2848.27 & 41.55  & \textbf{68.56 $\uparrow$}                  & 13   & 13    & 0              & 0.00\%           \\
		rand3reg\_70\_5 & 70 & 3317.86 & 51.91  & \textbf{63.92 $\uparrow$}                  & 13   & 14    & 1              & 7.69\%           \\
		rand3reg\_70\_6 & 70 & 4513.51 & 48.15  & \textbf{93.74 $\uparrow$}                  & 14   & 14    & 0              & 0.00\%           \\
		rand3reg\_70\_7 & 70 & 3490.30 & 37.27  & \textbf{93.65 $\uparrow$}                  & 13   & 13    & 0              & 0.00\%           \\
		rand3reg\_70\_8 & 70 & 3160.43 & 46.47  & \textbf{68.01 $\uparrow$}                  & 13   & 13    & 0              & 0.00\%           \\
		rand3reg\_70\_9 & 70 & 3753.55 & 47.61  & \textbf{78.83 $\uparrow$}                  & 15   & 14    & \textbf{-1}    & \textbf{-6.67\%} \\
		rand3reg\_80\_0 & 80 & TO      & 91.67  & \textbf{$\textgreater{}$109.09 $\uparrow$} & -    & 16    & -              & -                \\
		rand3reg\_80\_1 & 80 & TO      & 84.20  & \textbf{$\textgreater{}$118.76 $\uparrow$} & -    & 14    & -              & -                \\
		rand3reg\_80\_2 & 80 & TO      & 105.30 & \textbf{$\textgreater{}$94.96 $\uparrow$}  & -    & 16    & -              & -                \\
		rand3reg\_80\_3 & 80 & 9696.31 & 82.12  & \textbf{118.08 $\uparrow$}                 & 15   & 15    & 0              & 0.00\%           \\
		rand3reg\_80\_4 & 80 & 8088.37 & 73.03  & \textbf{110.76 $\uparrow$}                 & 13   & 14    & 1              & 7.69\%           \\
		rand3reg\_80\_5 & 80 & TO      & 78.26  & \textbf{$\textgreater{}$127.78 $\uparrow$} & -    & 14    & -              & -                \\
		rand3reg\_80\_6 & 80 & TO      & 92.69  & \textbf{$\textgreater{}$107.88 $\uparrow$} & -    & 14    & -              & -                \\
		rand3reg\_80\_7 & 80 & 8756.19 & 76.86  & \textbf{113.92 $\uparrow$}                 & 14   & 15    & 1              & 7.14\%           \\
		rand3reg\_80\_8 & 80 & 9334.07 & 67.08  & \textbf{139.16 $\uparrow$}                 & 14   & 13    & \textbf{-1}    & \textbf{-7.14\%} \\
		rand3reg\_80\_9 & 80 & 9630.43 & 72.94  & \textbf{132.03 $\uparrow$}                 & 14   & 14    & 0              & 0.00\%           \\
		rand3reg\_90\_0 & 90 & TO      & 161.38 & \textbf{$\textgreater{}$61.97 $\uparrow$}  & -    & 16    & -              & -                \\
		rand3reg\_90\_1 & 90 & TO      & 176.92 & \textbf{$\textgreater{}$56.52 $\uparrow$}  & -    & 17    & -              & -                \\
		rand3reg\_90\_2 & 90 & TO      & 161.42 & \textbf{$\textgreater{}$61.95 $\uparrow$}  & -    & 17    & -              & -                \\
		rand3reg\_90\_3 & 90 & TO      & 164.89 & \textbf{$\textgreater{}$60.65 $\uparrow$}  & -    & 16    & -              & -                \\
		rand3reg\_90\_4 & 90 & TO      & 170.63 & \textbf{$\textgreater{}$58.61 $\uparrow$}  & -    & 16    & -              & -                \\
		rand3reg\_90\_5 & 90 & TO      & 144.06 & \textbf{$\textgreater{}$69.42 $\uparrow$}  & -    & 16    & -              & -                \\
		rand3reg\_90\_6 & 90 & TO      & 162.69 & \textbf{$\textgreater{}$61.47 $\uparrow$}  & -    & 17    & -              & -                \\
		rand3reg\_90\_7 & 90 & TO      & 196.55 & \textbf{$\textgreater{}$50.88 $\uparrow$}  & -    & 17    & -              & -                \\
		rand3reg\_90\_8 & 90 & TO      & 179.23 & \textbf{$\textgreater{}$55.79 $\uparrow$}  & -    & 17    & -              & -                \\
		rand3reg\_90\_9 & 90 & TO      & 157.62 & \textbf{$\textgreater{}$63.44 $\uparrow$}  & -    & 17    & -              & -                \\
		&    & 3661.09 & 41.04  & \textbf{78.23 $\uparrow$}                  & 12.8 & 12.76 & \textbf{-0.04} & \textbf{-0.31\%} \\
		\bottomrule
	\end{tabular}
\end{table}

\begin{figure}[t]
	\centering
	\includegraphics[width=0.95\textwidth]{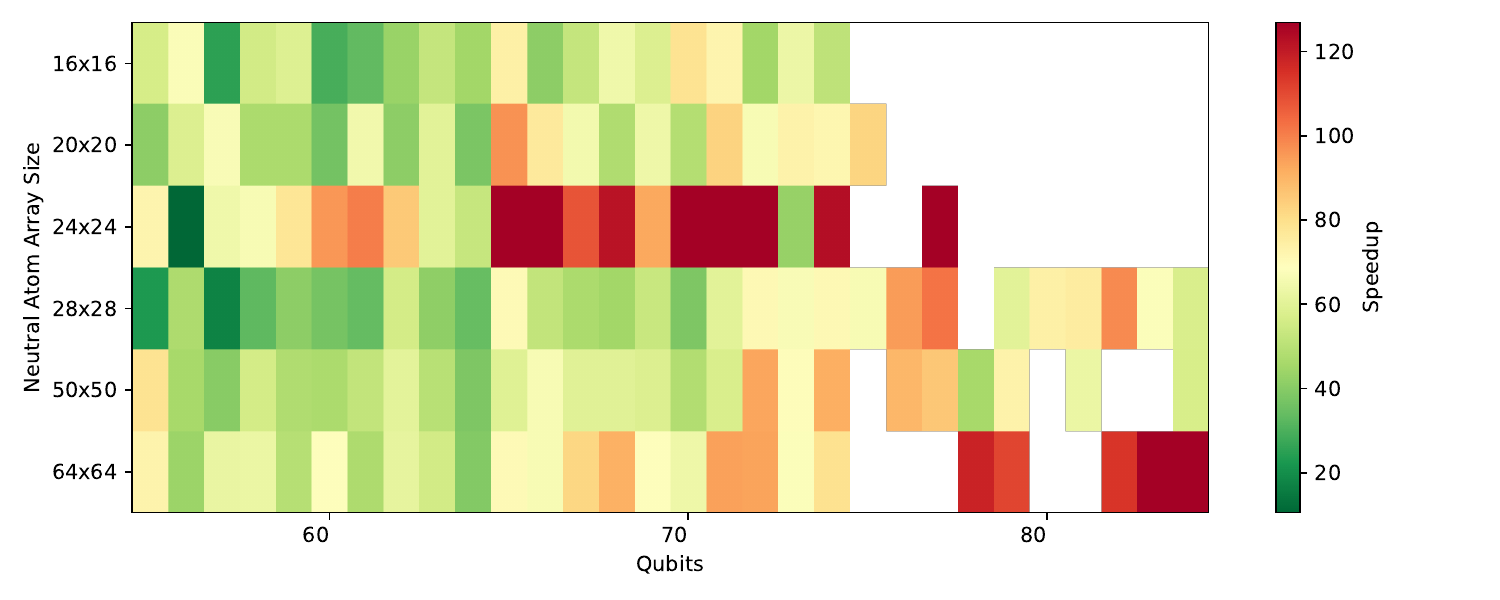}
	\caption{Heatmap visualization of PAC speedup factors relative to DPQA across varying qubit counts $(60-80)$ and neutral atom array dimensions. White cells indicate instances where DPQA exceeded the computation time threshold, while color intensity corresponds to speedup magnitude. Each qubit count category includes ten benchmark circuits include $rand3reg\_60\_0$ through $rand3reg\_60\_9$, $rand3reg\_70\_0$ through $rand3reg\_70\_9$, and $rand3reg\_80\_0$ through $rand3reg\_80\_9$.}
	\label{fig:heat}
\end{figure}

To better understand PAC's performance characteristics, Figure \ref{fig:heat} presents a heatmap illustrating the speedup of PAC over DPQA across different qubit counts and neutral atom array sizes.

The visualization shows that speedup factors generally increase with qubit count, indicating better relative performance on larger problem sizes. The $24\times24$ array configuration exhibits higher acceleration ratios compared to other array sizes for equivalent circuits.

\begin{figure}[h] %
	\centering
	\subfloat[]{\includegraphics[width=0.5\columnwidth, trim=1.5cm 0.5cm 1.5cm 0.5cm, clip]{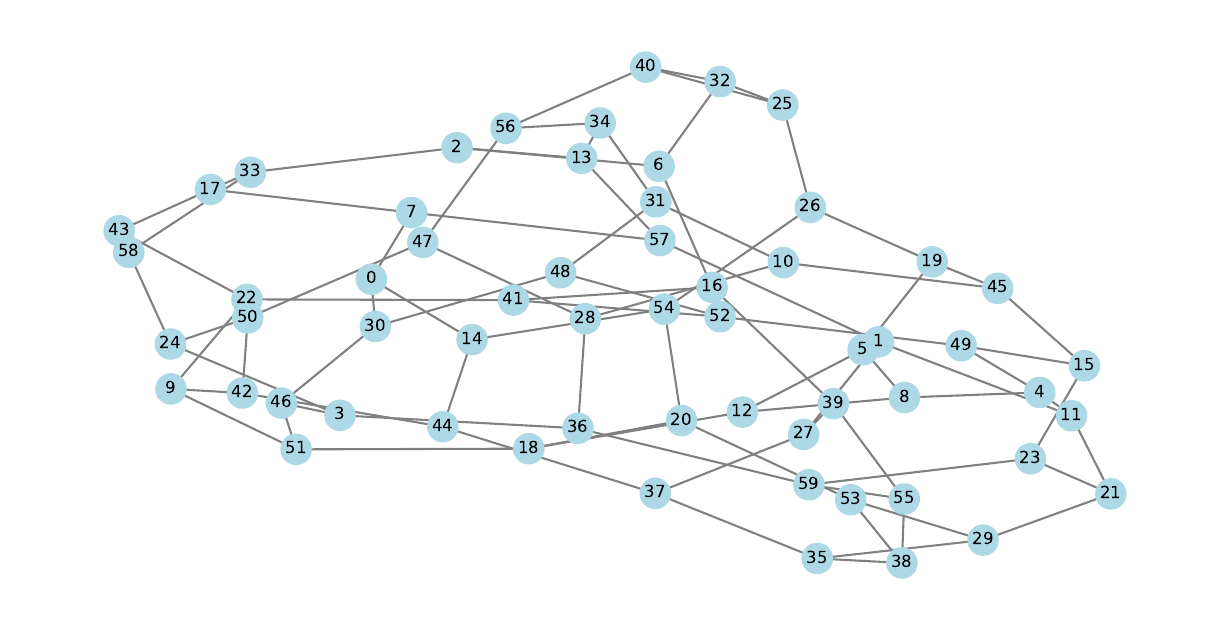}\label{fig:60-2}}
	\subfloat[]{\includegraphics[width=0.5\columnwidth, trim=1.5cm 0.5cm 1.5cm 0.5cm, clip]{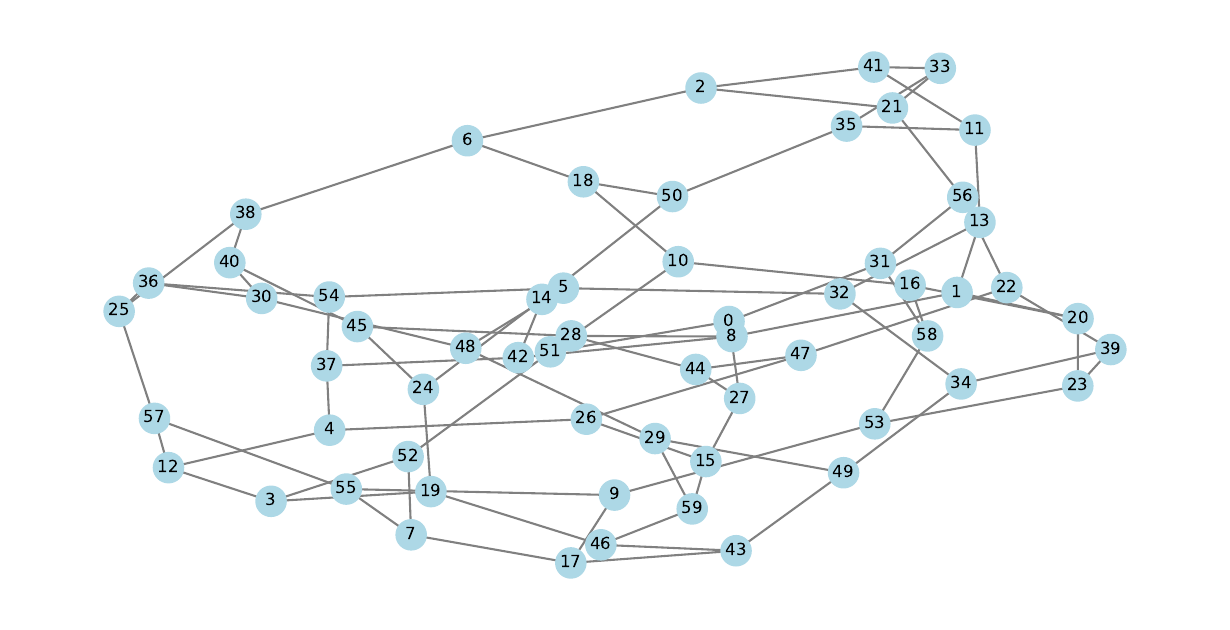}\label{fig:60-9}}
	\\
	\subfloat[]{\includegraphics[width=0.5\columnwidth, trim=1.5cm 0.6cm 2.5cm 1.8cm, clip]{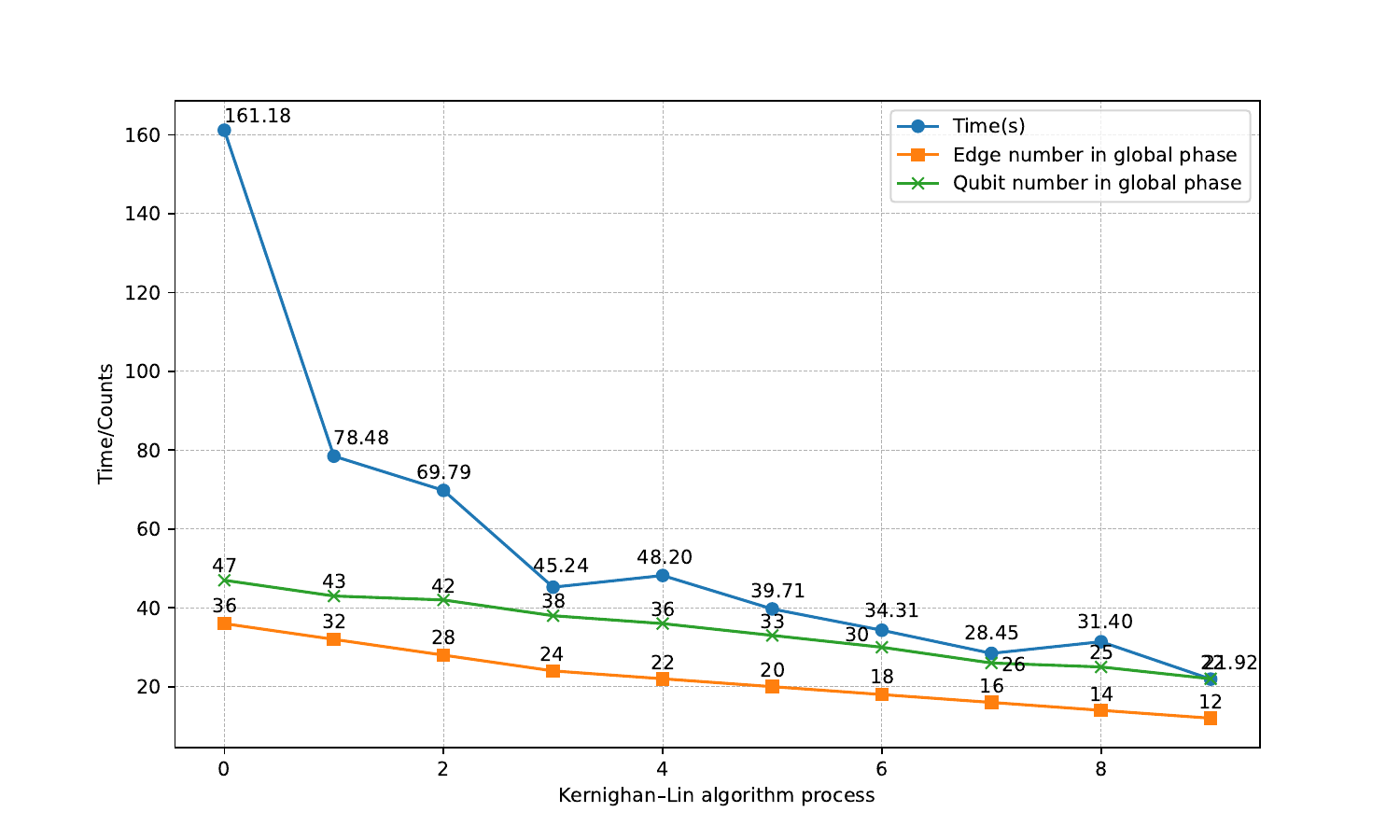}\label{fig:60-2-fig}}
	\subfloat[]{\includegraphics[width=0.5\columnwidth, trim=1.5cm 0.6cm 2.5cm 1.8cm, clip]{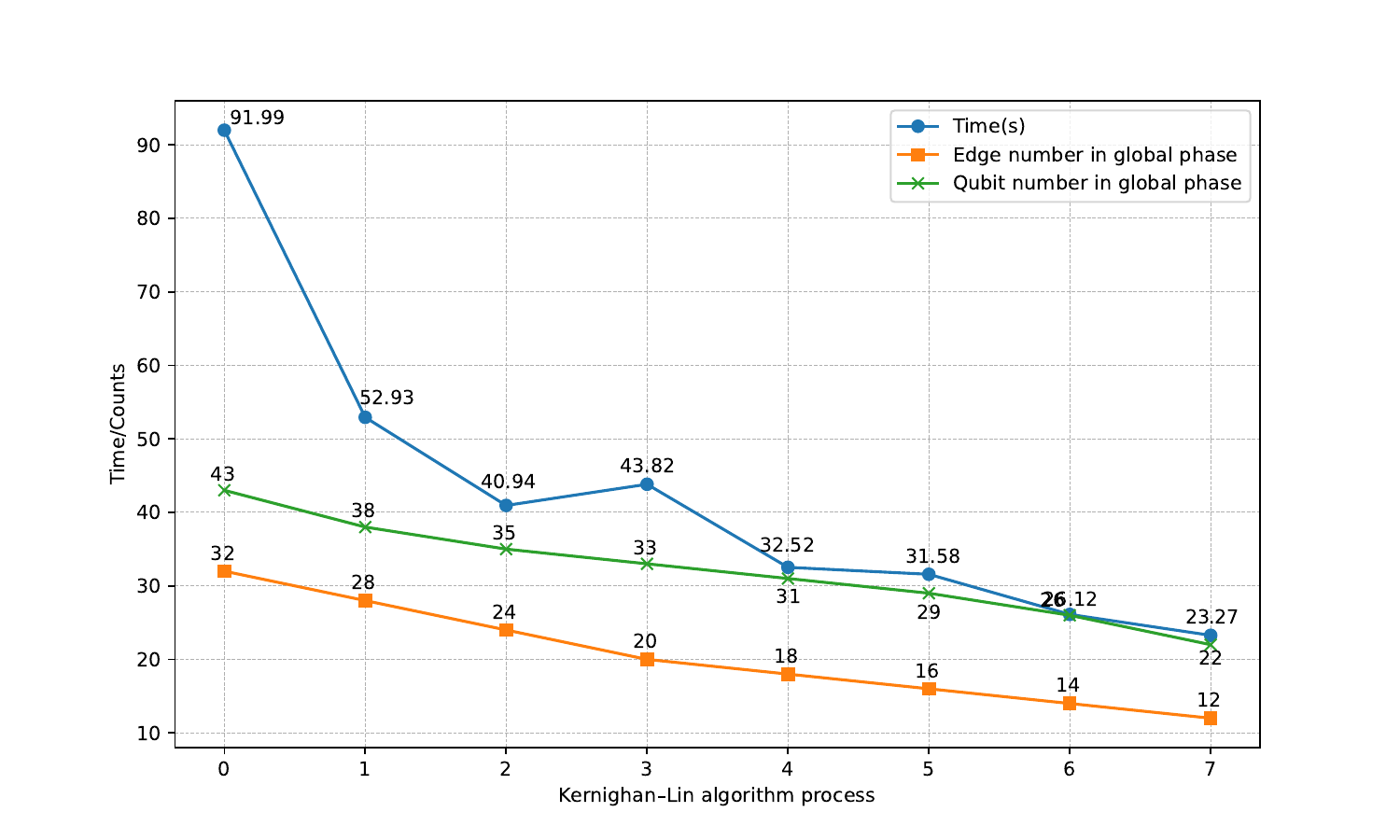}\label{fig:60-9-fig}}
	\caption{Ablation study of the Quantum Circuit Division process. (a) and (b) are two compiled 3-regular graph benchmarks, include $rand3reg\_60\_2$ of $60$ qubits and $rand3reg\_60\_9$ of $90$ edges. (c) and (d) represent the compilation metrics at progressive division stages for the corresponding benchmarks above. Blue lines represent compilation time, orange bars indicate the number of edges processed in the global phase, and blue-white bars show the number of qubits requiring global phase compilation.}
	\label{fig:ablation}
\end{figure}

The contribution of the quantum circuit division component can be observed through the ablation experiment in Figure \ref{fig:ablation}.

As the division process progresses, the number of edges between communities and the number of qubits requiring global phase compilation both decrease. This reduction in cross-region operations correlates with improved compilation efficiency, as shown by the downward trend in compilation time throughout the division process. The timing improvements confirm that the division strategy enhances local phase parallelism while reducing the computational load in the global phase.

\begin{figure}[h]
	\centering
	\includegraphics[width=0.55\columnwidth]{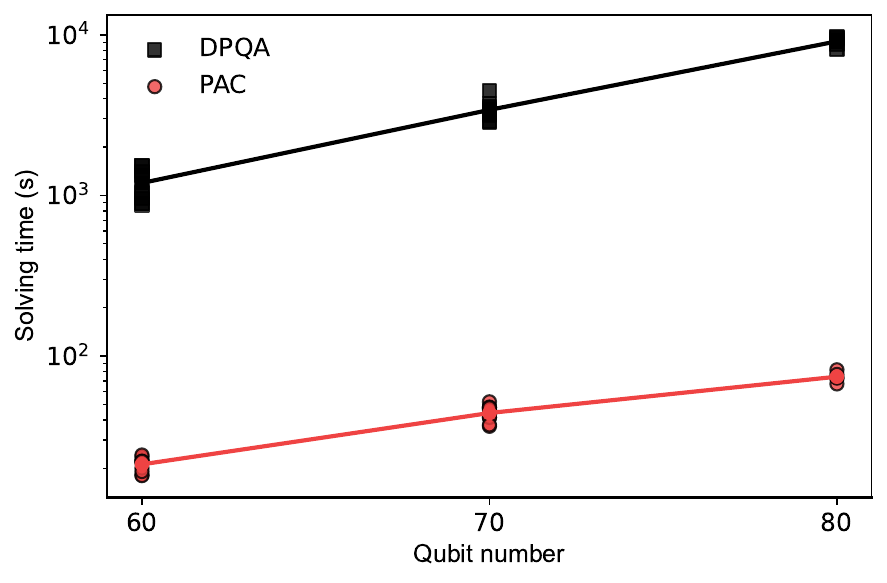}
	\caption{Comparison of compilation times between DPQA and PAC on a $64\times64$ neutral atom array, showing only circuits completed by DPQA within threshold. Data points represent individual circuits (DPQA: black squares; PAC: red circles), with logarithmic y-axis displaying performance scaling. Solid lines indicate mean compilation times per qubit count.}
	\label{fig:exp}
\end{figure}

Figure \ref{fig:exp} illustrates the different scaling behaviors of both methods with increasing problem sizes.

DPQA's solution time increases at a steeper rate with qubit count, while PAC shows a more moderate growth rate, especially for larger circuits. This difference in scaling patterns explains the widening performance gap observed in the earlier results and suggests that the physics-aware approach helps address efficiency limitations in conventional methods.

The analysis of these experimental results demonstrates that both Hardware Plane Partitioning and Quantum Circuit Division contribute to PAC's performance improvements. By incorporating physical hardware characteristics into the compilation process, PAC achieves efficiency gains while maintaining compilation quality, offering a promising approach for neutral atom quantum computer compilation as hardware scales to larger dimensions.

\section{Discussion}

Previous approaches to neutral atom quantum compilation have treated the problem primarily as computational, failing to effectively balance efficiency with hardware flexibility. PAC addresses this challenge by better leveraging the physical characteristics of quantum hardware, transforming the monolithic global problem into efficient local optimizations. This physics-aware approach achieved substantial performance improvements, with speedups of up to $78.5\times$ on $16\times16$ arrays and $139.16\times$ on $64\times64$ arrays compared to state-of-the-art methods, while maintaining comparable circuit quality. The increasing performance advantage at larger scales demonstrates PAC's scalability, making PAC valuable as quantum systems continue to grow.

The current implementation has limitations that represent common challenges in neutral atom compilation. The approach primarily uses equal hardware plane division that may not be optimal for all circuit topologies. Future research could develop circuit-aware partitioning strategies that dynamically adjust boundaries based on circuit topologies and qubit interaction patterns, particularly for irregular or asymmetric structures. Additionally, considerations of fidelity variations across the hardware plane could be incorporated into the optimization process. Dynamic hardware constraint modeling could further enhance compilation, potentially enabling even more efficient utilization of promising quantum computing platforms.

\section{Conclusion}

In this work, we propose PAC, a physics-aware compilation framework for neutral atom quantum computers that balances compilation efficiency with hardware flexibility. This approach integrates hardware physics characteristics into the compilation process, decomposing the global problem into local optimizations through hardware plane partitioning and quantum circuit division. Experiments demonstrate PAC's significant performance improvements across various array configurations while maintaining compilation quality.

Our exploration reveals two key insights into neutral atom quantum compilation. First, effectively exploiting the inherent parallelism of neutral atom architecture is crucial for improving compilation efficiency. The physical independence of different hardware regions creates natural opportunities for parallel processing that can be leveraged to reduce compilation time. Second, breaking down complex qubit mapping problems into simpler components through physics-guided partitioning transforms a challenging global optimization into more manageable sub-problems, improving efficiency while preserving solution quality.

Future work should focus on developing more adaptive physics-aware compilation strategies that can accommodate the diverse landscape of emerging quantum hardware architectures. Additionally, incorporating fidelity considerations into the compilation process would further leverage the unique advantages of neutral atom platforms, potentially enabling more robust quantum circuit execution as these systems scale toward practical quantum applications.

\section*{Authors' contributions}
\textbf{Geng Chen}: Writing - Original Draft, Visualization. \textbf{Guowu Yang}: Supervision, Formal analysis. \textbf{Wenjie Sun}: Methodology, Software. \textbf{Lianhui Yu}: Writing - Original Draft, Software. \textbf{Guangwei Deng}: Validation. \textbf{Desheng Zheng}: Methodology. \textbf{XiaoYu Li}: Conceptualization, Project administration.

\section*{Availability of supporting data}
All data generated or analysed during this study are available and included in this published article. \href{https://github.com/StillwaterQ/PAC}{Code} for our work has been open-sourced.

\section*{Acknowledgements}
This work was supported by National Natural Science Foundation of China (Grant No. 62472072, No. 62172075 and No. U2441217), Natural Science Foundation of Xinjiang Uygur Autonomous Region (No. 2024D01A20), National Key Research and Development Program of China (No. 2022YFA1405900).

\small
\bibliographystyle{elsarticle-num}
\bibliography{pacref} 

\newpage

\appendix 

\section{Quantum Circuit Basics}
\label{Quantum Circuit Basics}

Quantum computing is a computational paradigm that leverages quantum properties to perform calculations. It is typically implemented using quantum circuits, where the fundamental unit of computation is the qubit. By applying quantum gates to create entanglement and execute carefully designed operations, quantum circuits have the potential to solve a wide range of problems.

\begin{figure}[h]
	\centering
	\includegraphics[width=0.9\columnwidth, trim=7cm 8cm 7cm 8.5cm, clip]{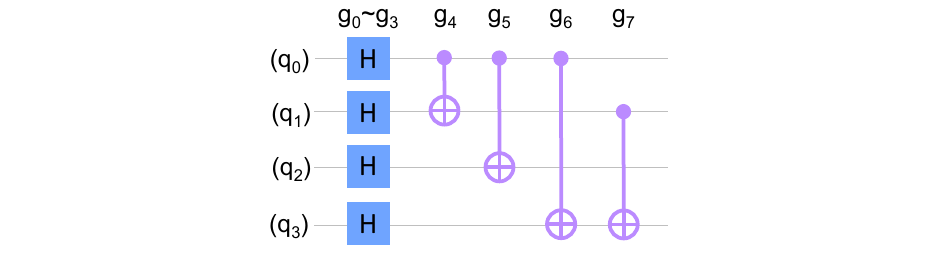}
	\caption{Demonstration of a quantum circuit.}
	\label{Fig.A1}
\end{figure}

Fig.~\ref{Fig.A1} illustrates a quantum circuit consisting of 4 qubits and 8 gates, where gates $g_0\sim g_3$ are single qubit gates and gates $g_4\sim g_7$ are two-qubit gates employed to establish entanglement between qubits. By performing measurement operations, we can obtain the output of a quantum circuit.

Similar to programs running on classical computers, quantum circuits must be deployed on quantum computers for execution. The compilation of quantum circuits involves transforming a quantum circuit into an executable sequence of operations that can be implemented on quantum hardware to achieve the intended functionality. The quantum compilation approach varies significantly depending on the underlying quantum hardware architecture,and this basic understanding of quantum circuits provides the foundation for appreciating the unique compilation challenges.

\section{Detailed Experimental Setup}
\label{Detailed Experimental Setup}

\textbf{Metrics} for evaluating compilation performance focused on both efficiency and quality. Compilation efficiency was assessed through solution time and speedup ratio measurements. Solution time records the total time required to complete the quantum circuit compilation process. A computation time limit of $10,000$ seconds was imposed on all compilation methods. If the compilation process exceeded this threshold, it was considered a timeout, and no valid solution was recorded. The speedup ratio was calculated as the ratio of solution times between DPQA and PAC, indicating how many times faster PAC completed the compilation compared to the baseline method. Compilation quality was evaluated using the number of quantum circuit layers produced after compilation. A smaller number of layers generally indicates higher quality compilation as it leads to shorter execution times on quantum hardware. The layer reduction ratio was calculated using Equation \ref{equ:layer-reduction-ratio}.

\textbf{Benchmarks} used in the evaluation consisted of $40$ random 3-regular graph circuits with qubit counts ranging from $60$ to $90$ and two-qubit gate counts from $90$ to $135$. These circuits were generated using a standard procedure for creating random regular graphs, where each vertex (qubit) has exactly three connections to other vertices. The circuits are structurally identical to those used in DPQA evaluation to ensure fair comparison. These 3-regular graph circuits were selected because they represent a wide range of quantum algorithms for combinatorial optimization problems. They are particularly relevant for applications such as Max-Cut optimization and multiple problem scenarios including Quantum Adiabatic Optimization Algorithm (QAOA) and trotterized adiabatic evolution. The structure of these circuits provides a realistic test case for evaluating compilation performance in practical quantum computing scenarios.

\textbf{Hardware} configurations for the neutral atom quantum computer included multiple array dimensions to comprehensively evaluate scalability. The configurations ranged from $16\times16$ arrays representing current proposed systems, through intermediate sizes of $20\times20$, $24\times24$, and $28\times28$, to large-scale configurations of $50\times50$ and $64\times64$ representing long-term scalability targets. These configurations were selected to evaluate performance across a spectrum of system sizes, from currently feasible implementations to potential future large-scale systems. This range allows for the assessment of how compilation performance scales with increasing system size, which is a critical factor for practical quantum computing applications.

\textbf{Environment} for all experiments utilized consistent hardware and software configurations to ensure fair comparison. Experiments were conducted on a Ryzen 9 7900X CPU running at 4.7GHz with 32GB of RAM, providing sufficient computational resources for evaluating compilation performance across various circuit and system sizes. The software environment consisted of Linux operating system, Python 3.9, Qiskit 1.0.0 for quantum circuit representation, NetworkX 3.2.1 for graph manipulation and analysis, and Z3 4.12.5.0 as the SMT solver for constraint satisfaction. The implementation of PAC utilized the same core solver mechanisms as DPQA, with the primary differences being the hardware plane partitioning and quantum circuit division components described in the main text.

\section{Experimentation on series array size}
\label{Experimentation on series array size}

Table \ref{tab:20x20} presents the results for a 20$\times$20 array. PAC completed all circuit compilations within 1000 seconds. The highest speedup was 96.56$\times$ on the benchmark rand3reg\_70\_0, and the average speedup across circuits compilable by DPQA reached 60.86$\times$. Regarding circuit depth, PAC outperformed DPQA with an average reduction of 0.43 layers.

\begin{table*}[h]
	\footnotesize
	\caption{Solving Time and Circuit Layers on 20$\times$20 Array.}
	\label{tab:20x20}
	\centering
	\tabcolsep=0.01\linewidth
	\renewcommand{\arraystretch}{1.0}
	\begin{tabular}{ccccccccc}
		\toprule
		\multirow{2}{*}{Quantum Circuit} & \multirow{2}{*}{Qubits} & \multicolumn{3}{c}{Solving time (s)}  & \multicolumn{4}{c}{Circuit Layers} \\
		\cmidrule(lr){3-5}\cmidrule(lr){6-9}
		&                         & DPQA & PAC      & Speedup & DPQA  & PAC  & $\Delta _{cl}$   & $R_{cl}$      \\
		\midrule
		rand3reg\_60\_0 & 60 & 1778.16 & 42.92  & \textbf{41.43 $\uparrow$}                 & 13 & 13       & 0                & 0.00\%            \\
		rand3reg\_60\_1 & 60 & 1891.21 & 32.43  & \textbf{58.31 $\uparrow$}                 & 12 & 12       & 0                & 0.00\%            \\
		rand3reg\_60\_2 & 60 & 2213.82 & 32.89  & \textbf{67.31 $\uparrow$}                 & 13 & 12       & \textbf{-1}      & \textbf{-7.69\%}  \\
		rand3reg\_60\_3 & 60 & 1659.24 & 35.22  & \textbf{47.12 $\uparrow$}                 & 12 & 12       & 0                & 0.00\%            \\
		rand3reg\_60\_4 & 60 & 1731.69 & 36.90  & \textbf{46.94 $\uparrow$}                 & 13 & 11       & \textbf{-2}      & \textbf{-15.38\%} \\
		rand3reg\_60\_5 & 60 & 1332.27 & 36.21  & \textbf{36.79 $\uparrow$}                 & 11 & 11       & 0                & 0.00\%            \\
		rand3reg\_60\_6 & 60 & 2016.17 & 31.10  & \textbf{64.83 $\uparrow$}                 & 12 & 11       & \textbf{-1}      & \textbf{-8.33\%}  \\
		rand3reg\_60\_7 & 60 & 1628.87 & 39.38  & \textbf{41.37 $\uparrow$}                 & 15 & 12       & \textbf{-3}      & \textbf{-20.00\%} \\
		rand3reg\_60\_8 & 60 & 2183.50 & 36.22  & \textbf{60.29 $\uparrow$}                 & 12 & 12       & 0                & 0.00\%            \\
		rand3reg\_60\_9 & 60 & 1442.68 & 37.77  & \textbf{38.19 $\uparrow$}                 & 11 & 12       & 1                & 9.09\%            \\
		rand3reg\_70\_0 & 70 & 6438.56 & 66.68  & \textbf{96.56 $\uparrow$}                 & 14 & 14       & 0                & 0.00\%            \\
		rand3reg\_70\_1 & 70 & 5241.48 & 68.91  & \textbf{76.07 $\uparrow$}                 & 14 & 14       & 0                & 0.00\%            \\
		rand3reg\_70\_2 & 70 & 5418.36 & 82.97  & \textbf{65.31 $\uparrow$}                 & 14 & 13       & \textbf{-1}      & \textbf{-7.14\%}  \\
		rand3reg\_70\_3 & 70 & 4024.76 & 84.12  & \textbf{47.85 $\uparrow$}                 & 13 & 13       & 0                & 0.00\%            \\
		rand3reg\_70\_4 & 70 & 5930.30 & 92.84  & \textbf{63.88 $\uparrow$}                 & 13 & 13       & 0                & 0.00\%            \\
		rand3reg\_70\_5 & 70 & 3882.56 & 79.52  & \textbf{48.82 $\uparrow$}                 & 13 & 12       & \textbf{-1}      & \textbf{-7.69\%}  \\
		rand3reg\_70\_6 & 70 & 5378.81 & 64.77  & \textbf{83.05 $\uparrow$}                 & 13 & 14       & 1                & 7.69\%            \\
		rand3reg\_70\_7 & 70 & 4280.42 & 64.02  & \textbf{66.86 $\uparrow$}                 & 15 & 13       & \textbf{-2}      & \textbf{-13.33\%} \\
		rand3reg\_70\_8 & 70 & 4359.31 & 59.55  & \textbf{73.21 $\uparrow$}                 & 13 & 12       & \textbf{-1}      & \textbf{-7.69\%}  \\
		rand3reg\_70\_9 & 70 & 4952.12 & 69.23  & \textbf{71.53 $\uparrow$}                 & 13 & 13       & 0                & 0.00\%            \\
		rand3reg\_80\_0 & 80 & 9166.24 & 111.32 & \textbf{82.34 $\uparrow$}                 & 14 & 15       & 1                & 7.14\%            \\
		rand3reg\_80\_1 & 80 & TO      & 125.70 & \textbf{$\textgreater{}$79.56 $\uparrow$} & -  & 15       & -                & -                 \\
		rand3reg\_80\_2 & 80 & TO      & 143.36 & \textbf{$\textgreater{}$69.76 $\uparrow$} & -  & 15       & -                & -                 \\
		rand3reg\_80\_3 & 80 & TO      & 144.07 & \textbf{$\textgreater{}$69.41 $\uparrow$} & -  & 16       & -                & -                 \\
		rand3reg\_80\_4 & 80 & TO      & 946.88 & \textbf{$\textgreater{}$10.56 $\uparrow$} & -  & 15       & -                & -                 \\
		rand3reg\_80\_5 & 80 & TO      & 139.32 & \textbf{$\textgreater{}$71.78 $\uparrow$} & -  & 15       & -                & -                 \\
		rand3reg\_80\_6 & 80 & TO      & 132.34 & \textbf{$\textgreater{}$75.56 $\uparrow$} & -  & 13       & -                & -                 \\
		rand3reg\_80\_7 & 80 & TO      & 148.79 & \textbf{$\textgreater{}$67.21 $\uparrow$} & -  & 16       & -                & -                 \\
		rand3reg\_80\_8 & 80 & TO      & 129.66 & \textbf{$\textgreater{}$77.13 $\uparrow$} & -  & 15       & -                & -                 \\
		rand3reg\_80\_9 & 80 & TO      & 217.81 & \textbf{$\textgreater{}$45.91 $\uparrow$} & -  & 13       & -                & -                 \\
		rand3reg\_90\_0 & 90 & TO      & 369.19 & \textbf{$\textgreater{}$27.09 $\uparrow$} & -  & 15       & -                & -                 \\
		rand3reg\_90\_1 & 90 & TO      & 243.11 & \textbf{$\textgreater{}$41.13 $\uparrow$} & -  & 17       & -                & -                 \\
		rand3reg\_90\_2 & 90 & TO      & 493.55 & \textbf{$\textgreater{}$20.26 $\uparrow$} & -  & 16       & -                & -                 \\
		rand3reg\_90\_3 & 90 & TO      & 857.46 & \textbf{$\textgreater{}$11.66 $\uparrow$} & -  & 15       & -                & -                 \\
		rand3reg\_90\_4 & 90 & TO      & 246.16 & \textbf{$\textgreater{}$40.62 $\uparrow$} & -  & 16       & -                & -                 \\
		rand3reg\_90\_5 & 90 & TO      & 221.91 & \textbf{$\textgreater{}$45.06 $\uparrow$} & -  & 16       & -                & -                 \\
		rand3reg\_90\_6 & 90 & TO      & 309.86 & \textbf{$\textgreater{}$32.27 $\uparrow$} & -  & 15       & -                & -                 \\
		rand3reg\_90\_7 & 90 & TO      & 302.84 & \textbf{$\textgreater{}$33.02 $\uparrow$} & -  & 16       & -                & -                 \\
		rand3reg\_90\_8 & 90 & TO      & 509.51 & \textbf{$\textgreater{}$19.63 $\uparrow$} & -  & 16       & -                & -                 \\
		rand3reg\_90\_9 & 90 & TO      & 228.26 & \textbf{$\textgreater{}$43.81 $\uparrow$} & -  & 16       & -                & -                 \\
		&    & 3664.31 & 57.38  & \textbf{60.86 $\uparrow$}                 & 13 & 12.57 & \textbf{-0.43} & \textbf{-3.30\%}   \\
		\bottomrule
	\end{tabular}
\end{table*}

As shown in Table \ref{tab:24x24}, PAC completed all compilations in under 300 seconds on the 24$\times$24 array. The maximum speedup reached 202.81$\times$ on rand3reg\_70\_1, and the average speedup on DPQA-solvable circuits was 102.56$\times$, implying PAC required less than 1\% of DPQA’s time on average. In terms of quality, PAC maintained comparable circuit depth, with DPQA and PAC averaging 12.48 and 12.67 layers, respectively.

\begin{table*}[h]
	\footnotesize
	\caption{Solving Time and Circuit Layers on 24$\times$24 Array.}
	\label{tab:24x24}
	\centering
	\tabcolsep=0.01\linewidth
	\renewcommand{\arraystretch}{1.0}
	\begin{tabular}{ccccccccc}
		\toprule
		\multirow{2}{*}{Quantum Circuit} & \multirow{2}{*}{Qubits} & \multicolumn{3}{c}{Solving time (s)}  & \multicolumn{4}{c}{Circuit Layers} \\
		\cmidrule(lr){3-5}\cmidrule(lr){6-9}
		&                         & DPQA & PAC      & Speedup & DPQA & PAC  & $\Delta _{cl}$   & $R_{cl}$      \\
		\midrule
		rand3reg\_60\_0 & 60 & 1550.36 & 21.52  & \textbf{72.03 $\uparrow$}                  & 11       & 12    & 1           & 9.09\%           \\
		rand3reg\_60\_1 & 60 & 1774.67 & 168.35 & \textbf{10.54 $\uparrow$}                  & 11       & 12    & 1           & 9.09\%           \\
		rand3reg\_60\_2 & 60 & 1618.20 & 25.06  & \textbf{64.56 $\uparrow$}                  & 12       & 12    & 0           & 0.00\%           \\
		rand3reg\_60\_3 & 60 & 2196.41 & 33.01  & \textbf{66.54 $\uparrow$}                  & 12       & 13    & 1           & 8.33\%           \\
		rand3reg\_60\_4 & 60 & 1985.55 & 25.57  & \textbf{77.64 $\uparrow$}                  & 13       & 13    & 0           & 0.00\%           \\
		rand3reg\_60\_5 & 60 & 2191.97 & 22.86  & \textbf{95.91 $\uparrow$}                  & 12       & 12    & 0           & 0.00\%           \\
		rand3reg\_60\_6 & 60 & 2038.84 & 20.37  & \textbf{100.11 $\uparrow$}                 & 12       & 11    & \textbf{-1} & \textbf{-8.33\%} \\
		rand3reg\_60\_7 & 60 & 1881.91 & 22.05  & \textbf{85.36 $\uparrow$}                  & 12       & 12    & 0           & 0.00\%           \\
		rand3reg\_60\_8 & 60 & 1367.69 & 22.66  & \textbf{60.37 $\uparrow$}                  & 11       & 12    & 1           & 9.09\%           \\
		rand3reg\_60\_9 & 60 & 1433.10 & 26.96  & \textbf{53.15 $\uparrow$}                  & 13       & 13    & 0           & 0.00\%           \\
		rand3reg\_70\_0 & 70 & 7590.54 & 42.97  & \textbf{176.66 $\uparrow$}                 & 12       & 13    & 1           & 8.33\%           \\
		rand3reg\_70\_1 & 70 & 6811.90 & 33.59  & \textbf{202.81 $\uparrow$}                 & 13       & 13    & 0           & 0.00\%           \\
		rand3reg\_70\_2 & 70 & 5301.34 & 49.09  & \textbf{107.98 $\uparrow$}                 & 12       & 13    & 1           & 8.33\%           \\
		rand3reg\_70\_3 & 70 & 4852.27 & 39.96  & \textbf{121.43 $\uparrow$}                 & 13       & 13    & 0           & 0.00\%           \\
		rand3reg\_70\_4 & 70 & 4262.68 & 46.16  & \textbf{92.35 $\uparrow$}                  & 12       & 13    & 1           & 8.33\%           \\
		rand3reg\_70\_5 & 70 & 5838.08 & 35.38  & \textbf{165.01 $\uparrow$}                 & 13       & 13    & 0           & 0.00\%           \\
		rand3reg\_70\_6 & 70 & 6825.88 & 50.74  & \textbf{134.53 $\uparrow$}                 & 15       & 14    & \textbf{-1} & \textbf{-6.67\%} \\
		rand3reg\_70\_7 & 70 & 6614.11 & 38.10  & \textbf{173.62 $\uparrow$}                 & 12       & 13    & 1           & 8.33\%           \\
		rand3reg\_70\_8 & 70 & 5175.00 & 120.03 & \textbf{43.11 $\uparrow$}                  & 14       & 14    & 0           & 0.00\%           \\
		rand3reg\_70\_9 & 70 & 4716.40 & 38.33  & \textbf{123.05 $\uparrow$}                 & 13       & 12    & \textbf{-1} & \textbf{-7.69\%} \\
		rand3reg\_80\_0 & 80 & TO      & 80.61  & \textbf{$\textgreater{}$124.06 $\uparrow$} & -        & 14    & -           & -                \\
		rand3reg\_80\_1 & 80 & TO      & 78.76  & \textbf{$\textgreater{}$126.97 $\uparrow$} & -        & 14    & -           & -                \\
		rand3reg\_80\_2 & 80 & 8185.80 & 64.42  & \textbf{127.08 $\uparrow$}                 & 14       & 13    & \textbf{-1} & \textbf{-7.14\%} \\
		rand3reg\_80\_3 & 80 & TO      & 78.24  & \textbf{$\textgreater{}$127.81 $\uparrow$} & -        & 13    & -           & -                \\
		rand3reg\_80\_4 & 80 & TO      & 92.88  & \textbf{$\textgreater{}$107.66 $\uparrow$} & -        & 14    & -           & -                \\
		rand3reg\_80\_5 & 80 & TO      & 103.84 & \textbf{$\textgreater{}$96.30 $\uparrow$}  & -        & 13    & -           & -                \\
		rand3reg\_80\_6 & 80 & TO      & 81.09  & \textbf{$\textgreater{}$123.32 $\uparrow$} & -        & 15    & -           & -                \\
		rand3reg\_80\_7 & 80 & TO      & 75.62  & \textbf{$\textgreater{}$132.25 $\uparrow$} & -        & 14    & -           & -                \\
		rand3reg\_80\_8 & 80 & TO      & 84.87  & \textbf{$\textgreater{}$117.83 $\uparrow$} & -        & 14    & -           & -                \\
		rand3reg\_80\_9 & 80 & TO      & 78.45  & \textbf{$\textgreater{}$127.47 $\uparrow$} & -        & 14    & -           & -                \\
		rand3reg\_90\_0 & 90 & TO      & 146.52 & \textbf{$\textgreater{}$68.25 $\uparrow$}  & -        & 17    & -           & -                \\
		rand3reg\_90\_1 & 90 & TO      & 143.11 & \textbf{$\textgreater{}$69.88 $\uparrow$}  & -        & 18    & -           & -                \\
		rand3reg\_90\_2 & 90 & TO      & 149.88 & \textbf{$\textgreater{}$66.72 $\uparrow$}  & -        & 17    & -           & -                \\
		rand3reg\_90\_3 & 90 & TO      & 178.82 & \textbf{$\textgreater{}$55.92 $\uparrow$}  & -        & 16    & -           & -                \\
		rand3reg\_90\_4 & 90 & TO      & 145.14 & \textbf{$\textgreater{}$68.90 $\uparrow$}  & -        & 17    & -           & -                \\
		rand3reg\_90\_5 & 90 & TO      & 214.41 & \textbf{$\textgreater{}$46.64 $\uparrow$}  & -        & 17    & -           & -                \\
		rand3reg\_90\_6 & 90 & TO      & 134.41 & \textbf{$\textgreater{}$74.40 $\uparrow$}  & -        & 16    & -           & -                \\
		rand3reg\_90\_7 & 90 & TO      & 163.04 & \textbf{$\textgreater{}$61.33 $\uparrow$}  & -        & 16    & -           & -                \\
		rand3reg\_90\_8 & 90 & TO      & 181.15 & \textbf{$\textgreater{}$55.20 $\uparrow$}  & -        & 17    & -           & -                \\
		rand3reg\_90\_9 & 90 & TO      & 137.61 & \textbf{$\textgreater{}$72.67 $\uparrow$}  & -        & 16    & -           & -                \\
		&    & 4010.13 & 82.39  & \textbf{96.94 $\uparrow$}                  & 12.47619 & 13.95 & 0.190476    & 1.53\%          \\
		\bottomrule
	\end{tabular}
\end{table*}

Table \ref{tab:28x28} reports results for a 28$\times$28 array. PAC achieved up to 112.84$\times$ speedup on rand3reg\_80\_3 and completed all circuits within 300 seconds. The average acceleration over DPQA was 56.54$\times$. PAC's average compiled circuit depth was 13.14, close to DPQA’s 12.76, demonstrating comparable quality across most circuits.

\begin{table*}[h]
	\footnotesize
	\caption{Solving Time and Circuit Layers on 28$\times$28 Array.}
	\label{tab:28x28}
	\centering
	\tabcolsep=0.01\linewidth
	\renewcommand{\arraystretch}{1.0}
	\begin{tabular}{ccccccccc}
		\toprule
		\multirow{2}{*}{Quantum Circuit} & \multirow{2}{*}{Qubits} & \multicolumn{3}{c}{Solving time (s)}  & \multicolumn{4}{c}{Circuit Layers} \\
		\cmidrule(lr){3-5}\cmidrule(lr){6-9}
		&                         & DPQA & PAC      & Speedup & DPQA & PAC  & $\Delta _{cl}$   & $R_{cl}$      \\
		\midrule
		rand3reg\_60\_0 & 60 & 786.15  & 33.84  & \textbf{23.23 $\uparrow$}                  & 10       & 11       & 1           & 10.00\%          \\
		rand3reg\_60\_1 & 60 & 1432.49 & 30.10  & \textbf{47.60 $\uparrow$}                  & 11       & 12       & 1           & 9.09\%           \\
		rand3reg\_60\_2 & 60 & 1430.77 & 80.86  & \textbf{17.69 $\uparrow$}                  & 14       & 14       & 0           & 0.00\%           \\
		rand3reg\_60\_3 & 60 & 1050.69 & 31.74  & \textbf{33.11 $\uparrow$}                  & 12       & 12       & 0           & 0.00\%           \\
		rand3reg\_60\_4 & 60 & 1365.81 & 33.16  & \textbf{41.19 $\uparrow$}                  & 11       & 12       & 1           & 9.09\%           \\
		rand3reg\_60\_5 & 60 & 1204.43 & 32.27  & \textbf{37.32 $\uparrow$}                  & 11       & 12       & 1           & 9.09\%           \\
		rand3reg\_60\_6 & 60 & 1016.73 & 29.97  & \textbf{33.93 $\uparrow$}                  & 12       & 12       & 0           & 0.00\%           \\
		rand3reg\_60\_7 & 60 & 1508.19 & 26.76  & \textbf{56.36 $\uparrow$}                  & 12       & 12       & 0           & 0.00\%           \\
		rand3reg\_60\_8 & 60 & 1089.72 & 26.02  & \textbf{41.88 $\uparrow$}                  & 12       & 11       & \textbf{-1} & \textbf{-8.33\%} \\
		rand3reg\_60\_9 & 60 & 1292.18 & 37.48  & \textbf{34.48 $\uparrow$}                  & 12       & 13       & 1           & 8.33\%           \\
		rand3reg\_70\_0 & 70 & 3367.48 & 47.75  & \textbf{70.52 $\uparrow$}                  & 13       & 14       & 1           & 7.69\%           \\
		rand3reg\_70\_1 & 70 & 3548.97 & 68.26  & \textbf{51.99 $\uparrow$}                  & 14       & 13       & \textbf{-1} & \textbf{-7.14\%} \\
		rand3reg\_70\_2 & 70 & 2672.21 & 56.76  & \textbf{47.08 $\uparrow$}                  & 12       & 13       & 1           & 8.33\%           \\
		rand3reg\_70\_3 & 70 & 2716.01 & 59.95  & \textbf{45.31 $\uparrow$}                  & 12       & 13       & 1           & 8.33\%           \\
		rand3reg\_70\_4 & 70 & 3246.66 & 60.65  & \textbf{53.53 $\uparrow$}                  & 13       & 13       & 0           & 0.00\%           \\
		rand3reg\_70\_5 & 70 & 2995.02 & 77.49  & \textbf{38.65 $\uparrow$}                  & 13       & 14       & 1           & 7.69\%           \\
		rand3reg\_70\_6 & 70 & 3165.55 & 52.51  & \textbf{60.28 $\uparrow$}                  & 14       & 14       & 0           & 0.00\%           \\
		rand3reg\_70\_7 & 70 & 2893.17 & 41.00  & \textbf{70.56 $\uparrow$}                  & 13       & 13       & 0           & 0.00\%           \\
		rand3reg\_70\_8 & 70 & 3170.81 & 47.11  & \textbf{67.30 $\uparrow$}                  & 13       & 13       & 0           & 0.00\%           \\
		rand3reg\_70\_9 & 70 & 2793.77 & 39.55  & \textbf{70.64 $\uparrow$}                  & 12       & 12       & 0           & 0.00\%           \\
		rand3reg\_80\_0 & 80 & 7376.53 & 110.48 & \textbf{66.77 $\uparrow$}                  & 13       & 14       & 1           & 7.69\%           \\
		rand3reg\_80\_1 & 80 & 8453.08 & 89.04  & \textbf{94.94 $\uparrow$}                  & 13       & 14       & 1           & 7.69\%           \\
		rand3reg\_80\_2 & 80 & 8623.65 & 84.29  & \textbf{102.31 $\uparrow$}                 & 13       & 13       & 0           & 0.00\%           \\
		rand3reg\_80\_3 & 80 & TO      & 88.62  & \textbf{$\textgreater{}$112.84 $\uparrow$} & -        & 14       & -           & -                \\
		rand3reg\_80\_4 & 80 & 6353.21 & 105.38 & \textbf{60.29 $\uparrow$}                  & 14       & 14       & 0           & 0.00\%           \\
		rand3reg\_80\_5 & 80 & 7241.10 & 97.74  & \textbf{74.08 $\uparrow$}                  & 15       & 15       & 0           & 0.00\%           \\
		rand3reg\_80\_6 & 80 & 7570.55 & 100.30 & \textbf{75.48 $\uparrow$}                  & 14       & 14       & 0           & 0.00\%           \\
		rand3reg\_80\_7 & 80 & 9909.30 & 101.30 & \textbf{97.82 $\uparrow$}                  & 15       & 15       & 0           & 0.00\%           \\
		rand3reg\_80\_8 & 80 & 7148.20 & 105.37 & \textbf{67.84 $\uparrow$}                  & 14       & 15       & 1           & 7.14\%           \\
		rand3reg\_80\_9 & 80 & 5914.57 & 103.10 & \textbf{57.37 $\uparrow$}                  & 13       & 14       & 1           & 7.69\%           \\
		rand3reg\_90\_0 & 90 & TO      & 168.48 & \textbf{$\textgreater{}$59.35 $\uparrow$}  & -        & 15       & -           & -                \\
		rand3reg\_90\_1 & 90 & TO      & 175.44 & \textbf{$\textgreater{}$57.00 $\uparrow$}  & -        & 16       & -           & -                \\
		rand3reg\_90\_2 & 90 & TO      & 187.09 & \textbf{$\textgreater{}$53.45 $\uparrow$}  & -        & 16       & -           & -                \\
		rand3reg\_90\_3 & 90 & TO      & 146.01 & \textbf{$\textgreater{}$68.49 $\uparrow$}  & -        & 16       & -           & -                \\
		rand3reg\_90\_4 & 90 & TO      & 128.03 & \textbf{$\textgreater{}$78.11 $\uparrow$}  & -        & 15       & -           & -                \\
		rand3reg\_90\_5 & 90 & TO      & 294.32 & \textbf{$\textgreater{}$33.98 $\uparrow$}  & -        & 16       & -           & -                \\
		rand3reg\_90\_6 & 90 & TO      & 181.32 & \textbf{$\textgreater{}$55.15 $\uparrow$}  & -        & 16       & -           & -                \\
		rand3reg\_90\_7 & 90 & TO      & 168.31 & \textbf{$\textgreater{}$59.41 $\uparrow$}  & -        & 16       & -           & -                \\
		rand3reg\_90\_8 & 90 & TO      & 233.97 & \textbf{$\textgreater{}$42.74 $\uparrow$}  & -        & 17       & -           & -                \\
		rand3reg\_90\_9 & 90 & TO      & 237.35 & \textbf{$\textgreater{}$42.13 $\uparrow$}  & -        & 17       & -           & -                \\
		&    & 3839.21 & 62.42  & \textbf{56.54 $\uparrow$}                  & 12.75862 & 13.13793 & 0.37931     & 2.97\%          \\
		\bottomrule
	\end{tabular}
\end{table*}

Table \ref{tab:50x50} shows the results for a 50$\times$50 array. PAC achieved a maximum speedup of 98.96$\times$ on rand3reg\_80\_5, with all compilation times remaining below 300 seconds. On circuits solvable by DPQA, the average speedup was 61.4$\times$. The average circuit depths were 12.96 for PAC and 12.77 for DPQA, indicating PAC’s significant efficiency gains without compromising output quality.

\begin{table*}[h]
	\footnotesize
	\caption{Solving Time and Circuit Layers on 50$\times$50 Array.}
	\label{tab:50x50}
	\centering
	\tabcolsep=0.01\linewidth
	\renewcommand{\arraystretch}{1.0}
	\begin{tabular}{ccccccccc}
		\toprule
		\multirow{2}{*}{Quantum Circuit} & \multirow{2}{*}{Qubits} & \multicolumn{3}{c}{Solving time (s)}  & \multicolumn{4}{c}{Circuit Layers} \\
		\cmidrule(lr){3-5}\cmidrule(lr){6-9}
		&                         & DPQA & PAC      & Speedup & DPQA & PAC  & $\Delta _{cl}$   & $R_{cl}$      \\
		\midrule
		rand3reg\_60\_0 & 60 & 1641.12 & 20.87  & \textbf{78.64 $\uparrow$}                 & 13       & 11       & \textbf{-2} & \textbf{-15.38\%} \\
		rand3reg\_60\_1 & 60 & 1336.53 & 28.83  & \textbf{46.37 $\uparrow$}                 & 12       & 13       & 1           & 8.33\%            \\
		rand3reg\_60\_2 & 60 & 1190.10 & 29.62  & \textbf{40.18 $\uparrow$}                 & 11       & 12       & 1           & 9.09\%            \\
		rand3reg\_60\_3 & 60 & 1951.21 & 34.64  & \textbf{56.32 $\uparrow$}                 & 13       & 13       & 0           & 0.00\%            \\
		rand3reg\_60\_4 & 60 & 1352.02 & 28.02  & \textbf{48.25 $\uparrow$}                 & 13       & 13       & 0           & 0.00\%            \\
		rand3reg\_60\_5 & 60 & 1481.97 & 31.44  & \textbf{47.13 $\uparrow$}                 & 12       & 12       & 0           & 0.00\%            \\
		rand3reg\_60\_6 & 60 & 1139.93 & 21.96  & \textbf{51.91 $\uparrow$}                 & 10       & 11       & 1           & 10.00\%           \\
		rand3reg\_60\_7 & 60 & 1425.03 & 23.44  & \textbf{60.79 $\uparrow$}                 & 11       & 11       & 0           & 0.00\%            \\
		rand3reg\_60\_8 & 60 & 1404.06 & 28.13  & \textbf{49.92 $\uparrow$}                 & 12       & 12       & 0           & 0.00\%            \\
		rand3reg\_60\_9 & 60 & 1173.01 & 30.52  & \textbf{38.43 $\uparrow$}                 & 12       & 13       & 1           & 8.33\%            \\
		rand3reg\_70\_0 & 70 & 3913.59 & 65.65  & \textbf{59.61 $\uparrow$}                 & 14       & 14       & 0           & 0.00\%            \\
		rand3reg\_70\_1 & 70 & 3314.47 & 49.84  & \textbf{66.50 $\uparrow$}                 & 13       & 12       & \textbf{-1} & \textbf{-7.69\%}  \\
		rand3reg\_70\_2 & 70 & 3127.76 & 52.40  & \textbf{59.69 $\uparrow$}                 & 12       & 13       & 1           & 8.33\%            \\
		rand3reg\_70\_3 & 70 & 2982.15 & 49.65  & \textbf{60.06 $\uparrow$}                 & 13       & 13       & 0           & 0.00\%            \\
		rand3reg\_70\_4 & 70 & 2329.54 & 39.94  & \textbf{58.32 $\uparrow$}                 & 11       & 12       & 1           & 9.09\%            \\
		rand3reg\_70\_5 & 70 & 2566.87 & 53.01  & \textbf{48.42 $\uparrow$}                 & 12       & 13       & 1           & 8.33\%            \\
		rand3reg\_70\_6 & 70 & 2702.81 & 47.09  & \textbf{57.40 $\uparrow$}                 & 12       & 13       & 1           & 8.33\%            \\
		rand3reg\_70\_7 & 70 & 4330.19 & 46.45  & \textbf{93.21 $\uparrow$}                 & 14       & 12       & \textbf{-2} & \textbf{-14.29\%} \\
		rand3reg\_70\_8 & 70 & 3272.70 & 47.07  & \textbf{69.53 $\uparrow$}                 & 13       & 14       & 1           & 7.69\%            \\
		rand3reg\_70\_9 & 70 & 3927.84 & 43.08  & \textbf{91.18 $\uparrow$}                 & 13       & 13       & 0           & 0.00\%            \\
		rand3reg\_80\_0 & 80 & TO      & 138.86 & \textbf{$\textgreater{}$72.02 $\uparrow$} & -        & 14       & -           & -                 \\
		rand3reg\_80\_1 & 80 & 7892.75 & 88.17  & \textbf{89.52 $\uparrow$}                 & 15       & 14       & \textbf{-1} & \textbf{-6.67\%}  \\
		rand3reg\_80\_2 & 80 & 7453.59 & 86.94  & \textbf{85.74 $\uparrow$}                 & 13       & 14       & 1           & 7.69\%            \\
		rand3reg\_80\_3 & 80 & 7102.36 & 153.01 & \textbf{46.42 $\uparrow$}                 & 15       & 15       & 0           & 0.00\%            \\
		rand3reg\_80\_4 & 80 & 8591.70 & 117.77 & \textbf{72.95 $\uparrow$}                 & 13       & 14       & 1           & 7.69\%            \\
		rand3reg\_80\_5 & 80 & TO      & 101.05 & \textbf{$\textgreater{}$98.96 $\uparrow$} & -        & 14       & -           & -                 \\
		rand3reg\_80\_6 & 80 & 7234.03 & 114.87 & \textbf{62.98 $\uparrow$}                 & 15       & 15       & 0           & 0.00\%            \\
		rand3reg\_80\_7 & 80 & TO      & 104.75 & \textbf{$\textgreater{}$95.47 $\uparrow$} & -        & 15       & -           & -                 \\
		rand3reg\_80\_8 & 80 & TO      & 110.31 & \textbf{$\textgreater{}$90.65 $\uparrow$} & -        & 15       & -           & -                 \\
		rand3reg\_80\_9 & 80 & 6831.75 & 119.99 & \textbf{56.93 $\uparrow$}                 & 15       & 15       & 0           & 0.00\%            \\
		rand3reg\_90\_0 & 90 & TO      & 209.22 & \textbf{$\textgreater{}$47.80 $\uparrow$} & -        & 16       & -           & -                 \\
		rand3reg\_90\_1 & 90 & TO      & 215.17 & \textbf{$\textgreater{}$46.48 $\uparrow$} & -        & 16       & -           & -                 \\
		rand3reg\_90\_2 & 90 & TO      & 158.44 & \textbf{$\textgreater{}$63.12 $\uparrow$} & -        & 16       & -           & -                 \\
		rand3reg\_90\_3 & 90 & TO      & 160.93 & \textbf{$\textgreater{}$62.14 $\uparrow$} & -        & 14       & -           & -                 \\
		rand3reg\_90\_4 & 90 & TO      & 176.66 & \textbf{$\textgreater{}$56.60 $\uparrow$} & -        & 16       & -           & -                 \\
		rand3reg\_90\_5 & 90 & TO      & 188.09 & \textbf{$\textgreater{}$53.17 $\uparrow$} & -        & 16       & -           & -                 \\
		rand3reg\_90\_6 & 90 & TO      & 233.19 & \textbf{$\textgreater{}$42.88 $\uparrow$} & -        & 16       & -           & -                 \\
		rand3reg\_90\_7 & 90 & TO      & 161.04 & \textbf{$\textgreater{}$62.10 $\uparrow$} & -        & 15       & -           & -                 \\
		rand3reg\_90\_8 & 90 & TO      & 183.16 & \textbf{$\textgreater{}$54.60 $\uparrow$} & -        & 16       & -           & -                 \\
		rand3reg\_90\_9 & 90 & TO      & 147.69 & \textbf{$\textgreater{}$67.71 $\uparrow$} & -        & 16       & -           & -                 \\
		&    & 3525.73 & 55.86  & \textbf{61.40 $\uparrow$}                 & 12.76923 & 12.96154 & 0.192308    & 1.51\%            \\
		\bottomrule
	\end{tabular}
\end{table*}

These results from intermediate array sizes consistently support the trend observed in the main text: PAC demonstrates substantial performance advantages across all tested architectures while maintaining compilation quality comparable to DPQA. The speedup factors range from approximately $50\times$ to over $200\times$ depending on the specific circuit and array configuration, with compilation times consistently remaining under $1200$ seconds even for the most complex circuits. This performance characteristic is particularly valuable for neutral atom quantum computing, where larger architectures are expected to become available as the technology matures.

\end{document}